\documentclass[12pt]{article}
\usepackage{jheppub,setspace,youngtab}

\newcommand{\EQ}[1]{\begin{align}\begin{split} #1 
\end{split}\end{align}}
\newcommand{\eq}[1]{\begin{equation}{ #1 
}\end{equation}}

\title{Central charges and RG flow of strongly-coupled $\mathcal{N}=2$ theory}
\author[\clubsuit]{Dan Xie} 

\author[\spadesuit]{and Peng Zhao}
\affiliation[\clubsuit]{Institute for Advanced Study, Einstein Dr., Princeton, NJ 08540, USA}

\affiliation[\spadesuit]{DAMTP, University of Cambridge, Cambridge CB3 0WA, UK}

\abstract{
We calculate the central charges $a, c$ and $k_G$ of a large class of four-dimensional $\mathcal{N} = 2$ superconformal field theories arising from compactifying the six-dimensional $\mathcal{N} = (2,0)$ theory on a Riemann surface with regular and irregular punctures. We also study the renormalization group flows between the general Argyres-Douglas theories, which all agree with the $a$-theorem.}
\begin{document}
\maketitle
\flushbottom
\section{Introduction}
In the past few years, a large class of four-dimensional $\mathcal{N}=2$ superconformal field theories (SCFTs) have been found using the magical
six-dimensional $A_{N-1}$ $(2,0)$ theory. These SCFTs fall into two classes: one class has integer scaling dimensions and dimensionless exactly marginal deformation, and the other class has fractional scaling dimensions\footnote{This class 
is actually much more general than people thought, and all the theories in the first class defined using the sphere have an irregular realization.}
and usually has dimensionful coupling constants. The first class is engineered using regular punctures (first-order poles) on a Riemann surface with arbitrary genus \cite{Gaiotto:2009we}, and the second class is engineered using a single irregular singularity (higher-order pole)
and at most one extra regular singularity on a sphere \cite{Xie:2012hs}. These SCFTs are generically strongly-coupled, namely one can not find a Lagrangian description in any duality frame.
The geometric construction leads to deep understanding about these theories, such as the three-dimensional mirror \cite{Benini:2010uu}, index \cite{Gadde:2011uv, Gaiotto:2012xa}, BPS spectrum, and wall-crossing \cite{Gaiotto:2009hg, Alim:2011kw, Gaiotto:2012db, Xie:2012gd}. In this paper, 
we will carry out a systematical calculation of the central charges for these new type of SCFTs.

Four-dimensional $\mathcal{N}=2$ theories can be characterized by several central charges. The central charges $a,c$ reflect the conformal anomaly, which is related to the $U(1)_R$ current anomaly 
due to the supersymmetry. The flavor central charge $k_G$ is defined using the two-point function of the global symmetry current, which is related to the three-point function of the $U(1)_R$ current
and two global symmetry currents. These central charges are independent of marginal deformations. So for the SCFT which has a free field description in the conformal manifold, one could easily compute these central charges. Such examples include the $\mathcal{N}=4$ super Yang-Mills theory and the $\mathcal{N}=2$ $SU(N)$
gauge theory with $2N$ flavors. 

\begin{figure}
\center
\includegraphics[width=14cm]{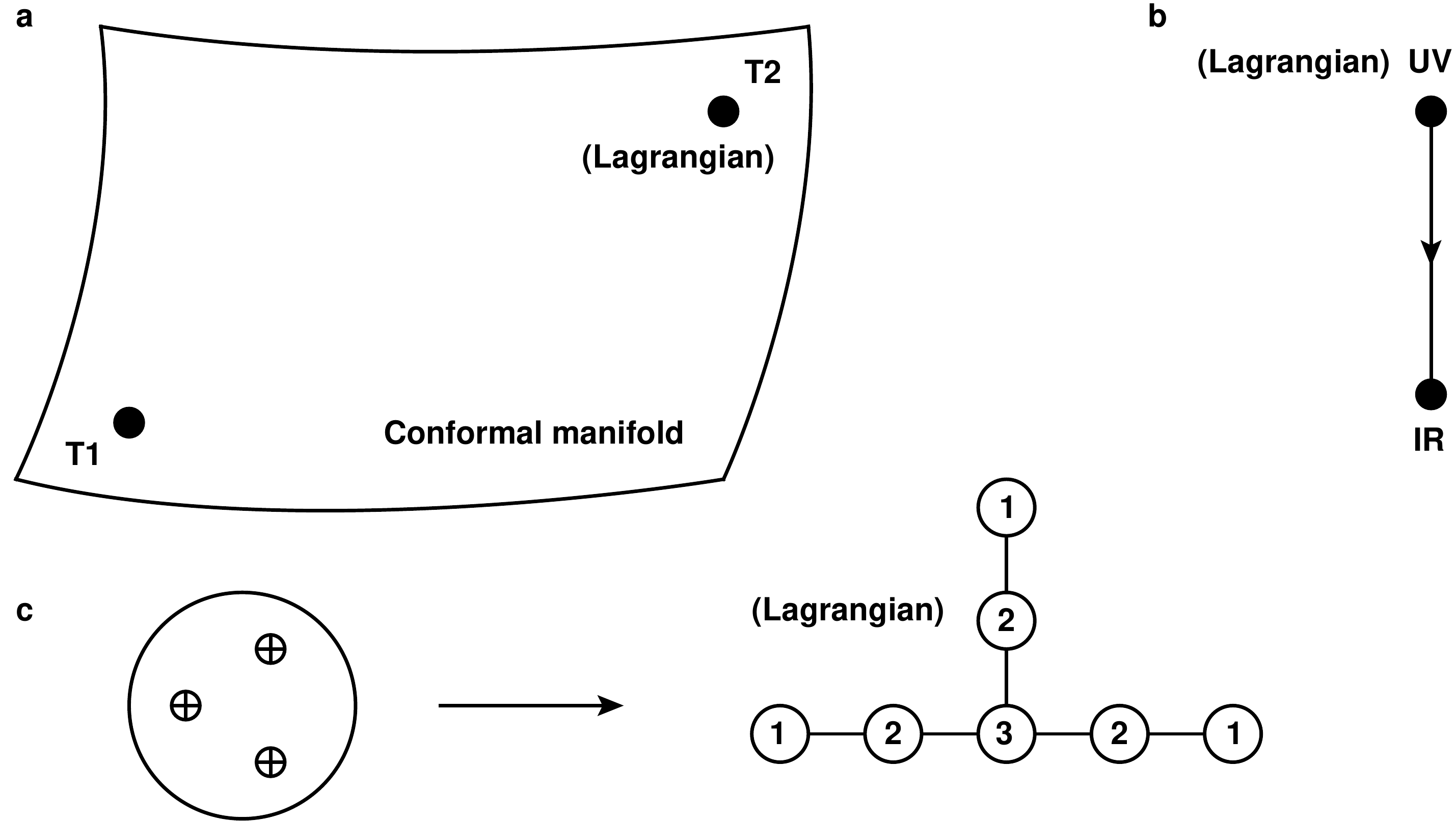}
\caption{The calculation of the central charges uses the weakly-coupled Lagrangian descriptions: a) $S$-duality, b) UV completion, c) 3d mirror.}
\label{intro}
\end{figure}

There are several methods one could use to study a strongly-coupled 4d SCFT, which all involve some kind of weakly-coupled description. 
The first method uses the highly nontrivial $S$-duality of the $\mathcal{N}=2$ field theory as pioneered by Argyres and Seiberg \cite{Argyres:2007cn}. The idea is the following: suppose we have a 
strongly-coupled field theory $A$ with non-abelian flavor symmetry, which one could gauge to form a weakly-coupled gauge theory description $T1$ with $A$ as 
the only strongly-coupled component. If there is a totally weakly-coupled description $T2$ in the $S$-duality frame whose central charges can be easily computed, then one could find the central charges of $A$ by subtracting the contribution from the weakly-coupled part of the original description. See figure \ref{intro}a.
A second method deals with the strongly-coupled theory where there is a nice weakly-coupled UV description. If we know the RG flow between a UV theory and 
the IR fixed point and no dramatic exotic description occurs, then the central charges can be found using the so-called $a$-maximization (the IR $U(1)_R$ symmetry
is a linear combination of the UV symmetry) \cite{Intriligator:2003jj}. See figure \ref{intro}b. The third method uses the supergravity dual in which one could do classical calculation to find the central charges \cite{Aharony:2007dj, Gaiotto:2009gz}.

However, for most of the strongly-coupled $\mathcal{N}=2$ theories such as the Argyres-Douglas theory, none of the above methods is applicable. There is another method
using the 3d mirror which always has a Lagrangian description. Using the information from the mirror theory, one could easily find the central charges by combining the 
scaling dimensions of the operator spectrum. See figure \ref{intro}c.

Quite remarkably, one can also calculate the central charges for those theories without the 3d mirror. We use a formula from \cite{Shapere:2008zf} which again relies on 
the IR Lagrangian description. It relates the central charges to the $R$-charges of topologically twisted theories via the 't Hooft anomaly matching condition. However, one of the $R$-charges $R(B)$ appearing in the formula is related to the discriminant of the Seiberg-Witten curve, which is difficult to compute. Nevertheless, it is possible to find $R(B)$ for these strongly-coupled theories, since they are part of a family of theories labeled by a discrete integer number $N$, and an infinite subset of this family has 3d mirror.
For theories that admit mirror pairs, we can compute the central charges directly and then infer $R(B)$ for this subset. We conjecture that it should extrapolate to all $N$. This is quite natural since $R(B)$ is related to the Seiberg-Witten curve which has a smooth form for this family. By using this method, we successfully calculate the central charges of all the SCFTs constructed using the punctured Riemann surface, and they recover all the calculations presented in the literature \cite{Aharony:2007dj, Shapere:2008zf, Gaiotto:2009gz, Chacaltana:2010ks, Xie:2012hs, Gaiotto:unpublished}.
Similarly, one can find the flavor central charge $k_G$ if the theory has non-abelian global symmetries.

For the Argyres-Douglas theory which has the relevant operators, one can turn on the relevant deformations and flow to other SCFTs in the class. We initiate a study of the RG flow between these theories and find that the central charge $a$ in the IR is smaller than the central charge in the UV. This confirms the the $a$-theorem \cite{Cardy:1988cwa, Jack:1990eb, Myers:2010tj, Komargodski:2011vj}, which states that the central charge $a$ decreases monotonically along the RG flow and thus measures the degrees of freedom at a given energy scale.

This paper is organized as follows. In section \ref{ac} we calculate the central charges $a$ and $c$ for all kinds of SCFTs mentioned above, and study the RG flow between the Argyres-Douglas theories to confirm the $a$-theorem. In section \ref{k} we compute the flavor central charge $k_G$. We conclude in section \ref{conclusion}. The derivation of the central charges from topological gauge theories will be reviewed in the appendix.

\section{Central charge $a,c$}\label{ac}
\subsection{Summary of calculation tools}
The central charges $a,c$ of a four-dimensional conformal field theory are defined using the operator product expansions of energy-momentum tensors. They 
can also be expressed as the coefficients of terms in the conformal anomaly \cite{Anselmi:1997am, Anselmi:1997ys, Kuzenko:1999pi}
\eq{
{\langle T^{\mu}_{\ \mu}\rangle = \frac{c}{16\pi^{2}} (\text{Weyl})^{2} - \frac{a}{16\pi^{2}} (\text{Euler})},
}
where the Weyl tensor and the Euler density associated with the background curvature are defined as
\EQ{
(\text{Weyl})^2&=R_{\mu\nu\rho\sigma}^2-2R_{\mu\nu}^2+{1\over 3}R^2, \\ (\text{Euler})&=R_{\mu\nu\rho\sigma}^2-4R_{\mu\nu}^2+R^2.
}
In supersymmetric theories, the central charges can be related to the $R$-charges, which are in turn related to the operator spectrum.
For weakly-coupled theories, one may easily find the central charges using the following formula \cite{Shapere:2008zf}
\eq{2a - c = \frac14 \sum_{i}\left(2[u_{i}] - 1 \right), \qquad a - c = \frac{n_{v}-n_{h}}{24} \label{Lagrangian},}
where we sum the scaling dimensions of all the Coulomb branch operators $u_{i}$. Here the difference between the number of vectormultiplets $n_{v}$ and the number of hypermultiplets $n_{h}$ is effectively equal to minus of the Higgs branch dimension, assuming that the gauge symmetry is completely broken. Consider for example the $SU(N)$ theory with $N_f=2N$. There are $N-1$ Coulomb branch operators with dimensions $2,\ldots, N$, and $n_v=N^2-1$ and $n_h=2N^2$. Using the above formula, the central charges are found as
\eq{
a=\frac{1}{24} \left(7 N^2-5\right),\qquad c=\frac{1}{6} \left(2 N^2-1\right).
}

For strongly-coupled theories, there is a formula from topological field theories \cite{Shapere:2008zf}, whose derivation will be reviewed in the appendix.
\eq{a = \frac{R(A)}4 + \frac{R(B)}6 + \frac{5r}{24}+ \frac{h}{24}, \qquad c = \frac{R(B)}3 + \frac{r}6 + \frac{h}{12}, \label{topological}} 
where $R(A)$ and $R(B)$ are the $R$-charges of the path integral measure factors; $r, h$ are respectively the number of free vectormultiplets and hypermultiplets at generic points of the Coulomb branch. For the type of theories we consider, $r$ coincides with the rank of the Coulomb branch, i.e. the number of operators whose scaling dimensions are larger than 1, and $h$ is zero. Using the familiar relation between the $R$-charges and the operator dimensions $R({\cal O})=2[{\cal O}]$, $R(A)$ can be expressed in terms of the scaling dimensions of the operator spectrum
\eq{
R(A) = \sum_{i}\left([u_{i}]-1\right).
}
$R(B)$ is determined from the discriminant of the Seiberg-Witten curve, which is in general very difficult to calculate. 

There is a special class of strongly-coupled field theories $I$ for which one could find the central charges using the $S$-duality. If the full theory has a Lagrangian description in 
another duality frame, then we could calculate the full central charges easily. The full theory in one duality frame is formed by gauging the flavor symmetry of $I$ and some other free matters; now the central charge of $I$ is derived by subtracting the contribution of the free matters. Let us look at the example of using Argyres-Seiberg duality to calculate the central charge of the $T_3$ theory. In one duality frame, the theory is $SU(3)$ with six flavors. In another duality frame, the theory is an $SU(2)$ group coupled with one fundamental and the $T_3$ theory. The central charges using the $SU(3)$ duality frame is $a=58/24,~c=17/6$. In the other duality frame, the contribution of the $SU(2)$ with one fundamental is $a=17/24,~c=4/6$. Therefore the 
central charge of the $T_3$ theory is 
\eq{
a=\frac{41}{24}, \qquad c=\frac{13}{6}.
}
For the general strongly-coupled theory $I$, it is hard to find such good full theory which satisfy the above two constraints: $I$ is a component in one duality frame and the full
theory has a weakly-coupled description in another duality frame. 

\begin{figure}
\center
\includegraphics[scale=0.4]{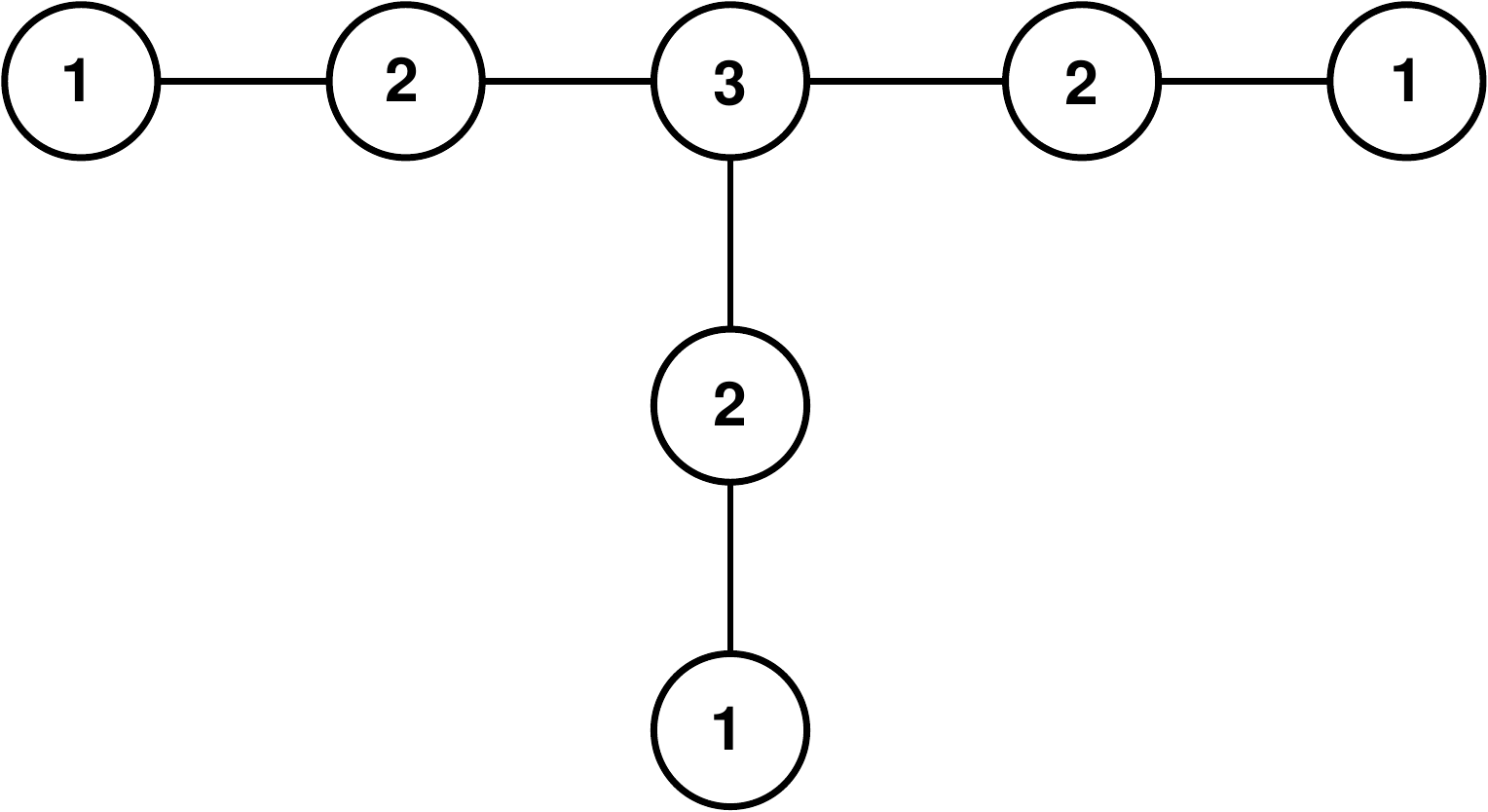}
\caption{3d mirror for the $T_{3}$ theory}
\label{T3}
\end{figure}

In this paper, we take another approach by using the formula (\ref{Lagrangian}) to calculate the central charges of a large class of SCFTs engineered using the six-dimensional $(2,0)$ theory. By looking at that formula, we can regard $n_h-n_v$ as the dimension of 
the Higgs branch, and such number could be read from the three-dimensional mirror which usually has a Lagrangian description! By this method, one can easily find the central charges of many theories. Let us illustrate the above idea for the $T_3$ theory. The three-dimensional mirror is shown in figure \ref{T3}, whose Coulomb branch
dimension is 11 which gives the Higgs branch dimension of the $T_3$ theory. The $T_3$ theory has one Coulomb branch operator with dimension 3. Substituting the above
information into the formula (\ref{Lagrangian}), we have the equations
\eq{
2a-c={5\over4},\qquad a-c=-{11\over24}.
}
Hence $a={41\over24}$ and $c={13\over6 }$, which matches the result from using the $S$-duality.

However, there are also strongly-coupled Argyres-Douglas theories which do not admit three-dimensional mirrors. For these theories, our strategy for calculating the central charges is to assume that $R(B)$ has a universal form for the theories in a family labeled by an integer $N$, i.e. it is a function of $N$. In this case the formula (\ref{topological}) turns out to be 
quite useful. For the subset of theories which have 3d mirrors, we can compute $R(B)$ using the explicit central charges found from the formula
(\ref{Lagrangian}). 
After finding out $R(B)$, we could also calculate the central charges of the strongly-coupled theories without 3d mirror.

\subsection{Regular punctures}
A large class of $\mathcal{N}=2$ superconformal field theories can be engineered by compactifying the six-dimensional $A_{k-1}$ $(2,0)$ theory on a Riemann surface with regular punctures (first-order poles).\footnote{This type includes the $A_{N}$ and the affine $A_N$ type quiver theories with Lagrangian description considered in \cite{Witten:1997sc}. The $A_N$ quiver is realized as a sphere with several simple punctures and two generic punctures, and the affine $A_N$ quiver is realized as a torus with several simple punctures. It also includes the strongly-coupled theory with $E_n$ flavor symmetry considered in \cite{Minahan:1996fg, Minahan:1996cj}, i.e. the $E_6$ theory is realized as the $A_2$ theory on a sphere with three full punctures.} The data in defining the theories are
\begin{enumerate}
\item[a.] A punctured Riemann surface with genus $g$.
\item[b.] Each puncture is labeled by a Young tableau $Y=[h_{1},h_2,\ldots, h_r]$ with a total of $k=\sum h_i$ boxes.\end{enumerate}
The gauge coupling constants are identified with the complex structure moduli of the Riemann surface. The mass parameters and flavor symmetry can be read from the Young tableau, i.e. there are $r-1$ mass parameters and the flavor symmetry from each 
puncture is 
\eq{
S\left[\prod_{\l_{h}>0} U(l_h)\right],
}
where $l_h$ is the number of columns with height $h$ in the Young tableau. The full flavor symmetry is usually the direct sum of the flavor symmetries from all the punctures (sometimes there is an enhancement). The $S$-duality is realized as different degeneration limits of the same Riemann surface \cite{Gaiotto:2009we}. The weakly-coupled gauge group and the matter content in each duality frame are solved in \cite{Nanopoulos:2010ga}, see also \cite{Chacaltana:2010ks} for some examples.

The number of Coulomb branch operators can be counted using the data in each Young tableau.
Let us focus on the sphere case by taking $g=0$ (the higher genus case is simple once the sphere case is understood). 
Define $p^{(j)}_{i}=i-s^{(j)}_i$, where $s^{(j)}_i$ is the height of the $i$-th box in the $j$-th tableau. The number of dimension $i$ ($2\leq i\leq k$) operators of the SCFT can be obtained by summing over all the Young tableaux
\eq{
d_i=\sum_j p_{i}^{(j)}-2i+1.
\label{dimension}
}
Thus the operator spectrum can be easily determined from the geometric data. The IR behavior of the Coulomb branch can also be found from the geometric construction. The Seiberg-Witten curve corresponds to the spectral curve of the Hitchin integrable system defined by the above data:
\eq{
x^k+\sum_{i=2}^k\Phi_i(z)x^{k-i}=0,
}
where $z$ is the coordinate on the Riemann surface and $x$ is the coordinate on the cotangent bundle;
$\Phi_{i}$ is a degree $i$ meromorphic differential on the Riemann surface whose pole structure at the singularity is determined by Young tableaux. 

To use the formula (\ref{Lagrangian}), we need to know the effective number of the Higgs branch dimension. This can be read from the 3d mirror \cite{Intriligator:1996ex, deBoer:1996ck, Gaiotto:2008ak} as follows. We compactify the 4d theory on a circle and flow in the deep IR to a 3d $\mathcal{N}=4$ SCFT. Mirror symmetry exchanges the Coulomb branch of this theory and the Higgs branch of the mirror theory, and vice versa. The wonderful thing
 is that the 3d mirror for the class of theories considered above always has a Lagrangian description. Moreover, the mirror theory is always ``good''\footnote{A 3d ``good'' theory
 means that the $R$-symmetry in the IR is the same as the UV theory, as defined in \cite{Gaiotto:2008ak}. For the star-shaped quiver considered here, ``good" theory
 means that $N_c\geq2N_f$ for all the quiver nodes.} considering only the non-degenerate case when there are sufficiently many punctures.
So the knowledge of the Coulomb branch of the mirror theory would give us the Higgs branch dimension of the original theory.

\begin{figure}
\center
\includegraphics[scale=0.4]{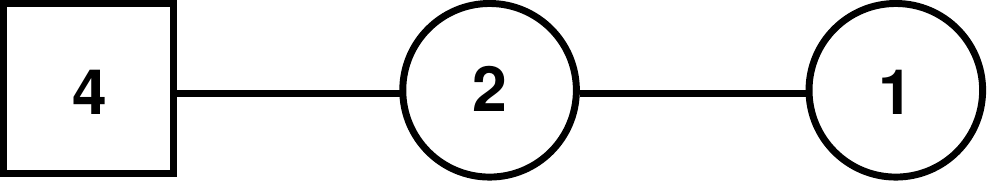}
\caption{Quiver tail for a regular puncture in the mirror theory.}
\label{reg}
\end{figure}

Let us review how to obtain the mirror theory of a given $\mathcal{N}=2$ theory defined on a sphere with regular punctures. Each regular puncture is associated with a Young tableau $Y_{0}$, and one can associate a quiver tail with it. Suppose the tableau has height $Y_{0} = [h_{1}, h_{2}, \ldots, h_{r}]$, then we get in the mirror theory a quiver of the form 
\eq{\boxed{k}-SU(h_{2}+\cdots+h_{r})-\cdots-SU(h_{r-1}+h_{r})-SU(h_{r}),}
where the leftmost box $\boxed{k}$ denotes the global flavor symmetry group $SU(k)$. For example, the quiver tail corresponding to the tableau $\tiny\yng(1,3)$ is shown in figure \ref{reg}.

\begin{figure}
\center
\includegraphics[width=15cm]{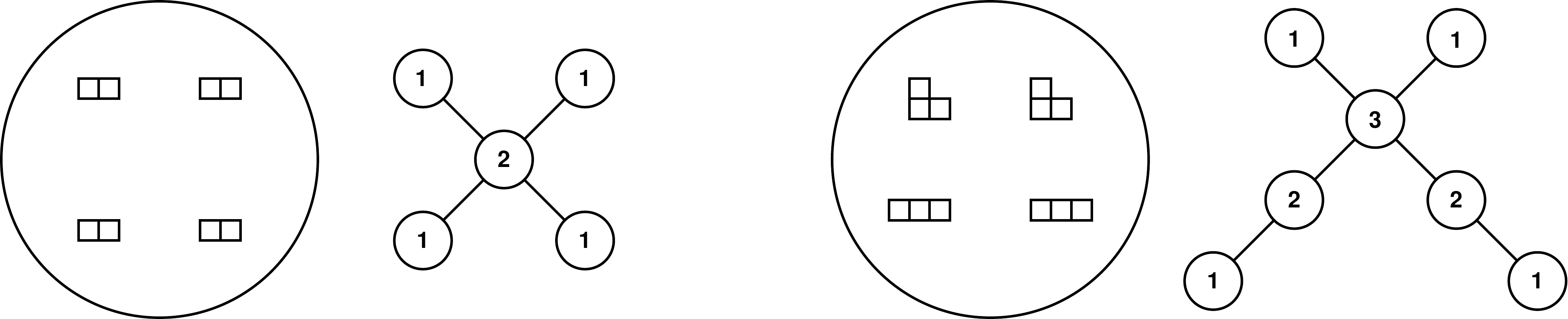} 
\caption{Gaiotto construction of superconformal QCD with $N_{f} = 4$ and $N_{f} = 6$ on a sphere with four punctures, and their mirror pairs. One can find that the true flavor symmetry is $U(2N_f)$ and $U(1)$, since there is a 
chain of $A_{2N_f}$ subquiver which satisfies the 4d $\mathcal{N}=2$ conformal relation, which will give the $U(2N_f)$ symmetry.}
\label{mirror}
\end{figure}

The full 3d mirror is derived by gauging the diagonal $SU(k)$ symmetries of all the quiver tails. It is a star-shaped quiver \cite{Benini:2010uu}, which always has 
a Lagrangian description. Let us consider two simple examples: the $SU(2)$ theory with $N_f=4$ and the $SU(3)$ theory with $N_f=6$. The six-dimensional construction and the mirror quiver are shown in figure \ref{mirror}.\footnote{One can actually see the enhancement of the flavor symmetry using the result in \cite{Gaiotto:2008ak}: if there is a subquiver with the $ADE$ shape and each quiver node satisfies the condition $N_f=2N_c$, then there is an enhanced flavor symmetry of the corresponding $ADE$ type.}
The Higgs branch dimension of the original theory is equal to the Coulomb branch dimension of the star-shaped quiver. 

With the knowledge of the operator spectrum and the 3d mirror, one can easily find out the central charges of any SCFT defined using regular punctures on a sphere.
Let us give a simple example here. 

\begin{figure}
\center
\includegraphics[scale=0.4]{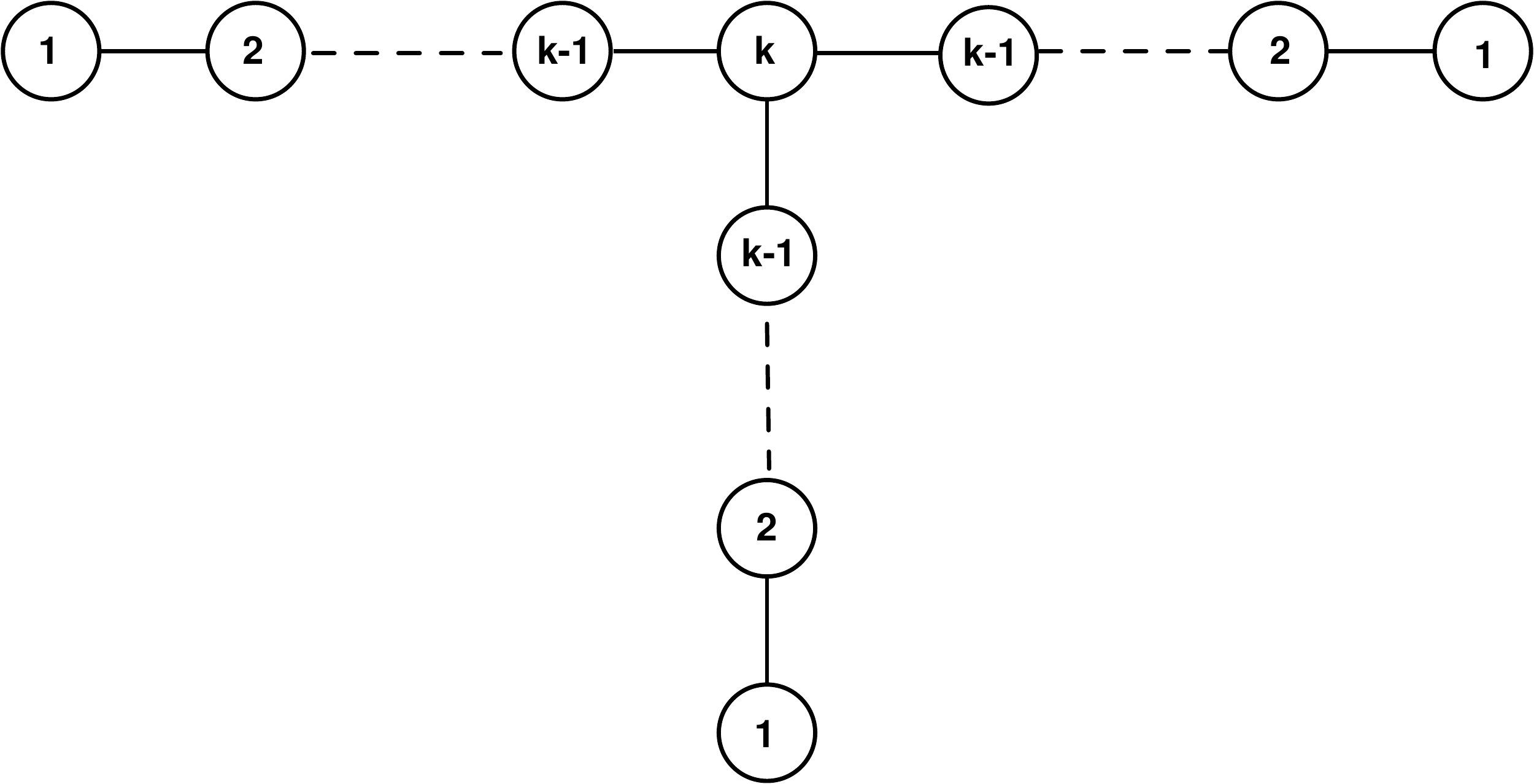}
\caption{3d mirror for the $T_{k}$ theory.}
\label{TN}
\end{figure}

\paragraph{Example 1:} The $T_k$ theory is realized as a sphere with three full punctures, which are each labelled by the Young tableau $Y = [1, \ldots, 1]$. Since each column has height 1, we have $p^{(j)}_{i} = i-1$. The number of dimension $i$ operators is
\eq{
d_i=\sum^{3}_{j=1}p^{(j)}_{i} - 2i + 1 = i-2.
}
The Higgs branch dimension read from the Coulomb branch dimension of the mirror (see figure \ref{TN}) is
\eq{
-(n_v-n_h)={1\over2}(k-1)(3k+2).
}

Using the formula (\ref{Lagrangian}), we have
\EQ{
2a-c&={1\over 4}\sum_{i=2}^k(2i-1)(i-2)=\frac{1}{24} (k-2) (k-1) (4 k+3),	 \\
a-c&=-\frac{1}{48} (k-1) (3 k+2).
}
Hence the central charges are
\eq{
a_{T_k}=\frac{k^3}{6}-\frac{5 k^2}{16}-\frac{k}{16}+\frac{5}{24}, \qquad c_{T_k}=\frac{k^3}{6}-\frac{k^2}{4}-\frac{k}{12}+\frac{1}{6},
\label{T_k}
}
which agrees with the result in the literature \cite{Gaiotto:2009gz}. Using (\ref{topological}), we can calculate $R(B)$
\eq{R(B) = \frac{1}{2} k(k-1)^2.\label{RBTk}}
Although we do not use this function to calculate the central charges, it seems to have interesting connection to the number of BPS states of this theory.

For theories defined using the higher genus Riemann surface, the Coulomb branch dimension of the mirror does not give 
the right number of the Higgs branch dimension of the original theory (something exotic happens in the IR). We are going to use a result found in \cite{Nanopoulos:2010ga}: in degenerating the handles, one always has an $SU(k)$ gauge group, and there are two more full punctures in the complete degeneration limit.
So one can always degenerate the higher genus theory to a sphere with several 
newly appeared full punctures. As we described earlier, one can find the central charges of the sphere part, and then add the 
contribution from each decoupled gauge group. The central charge contribution from the decoupled gauge group part is (the Coulomb branch operator dimensions are $2,\ldots,k$ and $n_v=k^2-1$): 
\eq{
a_{SU(k)}=\frac{5}{24} \left(k^2-1\right),\qquad c_{SU(k)}=\frac{1}{6} \left(k^2-1\right). 
\label{SUk}
}

\begin{figure}
\center
\includegraphics[scale=0.3]{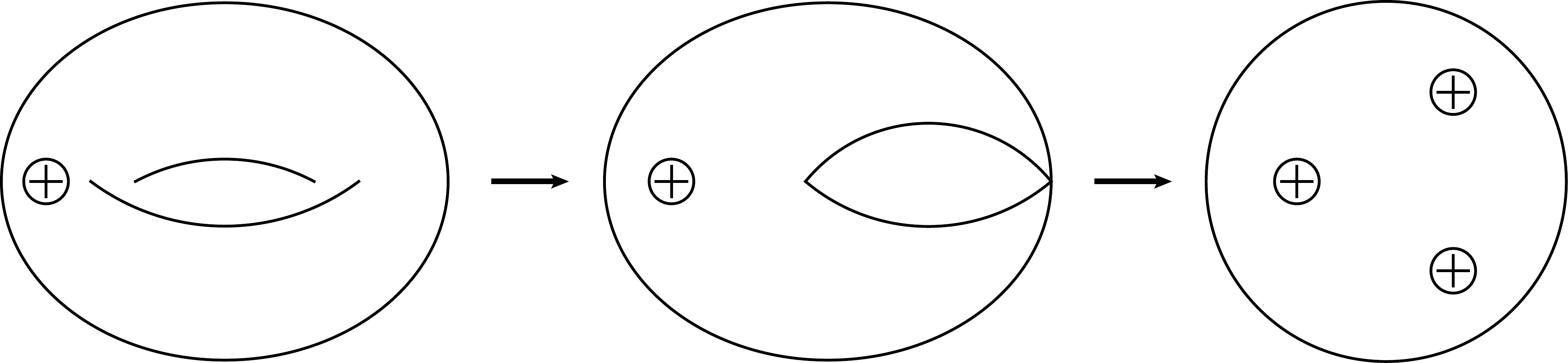}
\caption{Degeneration of a torus with one puncture into a sphere with three punctures. Circles with cross denote full punctures. At the degeneration point, we obtain two new full punctures.}
\label{degeneration}
\end{figure}

\paragraph{Example 2:} Consider a theory realized by a torus with one full puncture. We get the $T_k$ theory in the complete degeneration limit (see figure \ref{degeneration}),
plus a decoupled $SU(k)$ group. 
\noindent
Combining the central charges for each component (\ref{T_k}) and (\ref{SUk}), one has
\EQ{
a&=a_{T_k}+a_{SU(k)}=\frac{k^3}{6}-\frac{5 k^2}{48}-\frac{k}{16},\\
c&=c_{T_k}+c_{SU(k)}=\frac{k^3}{6}-\frac{k^2}{12}-\frac{k}{12}.
}

\subsection{Irregular punctures: general Argyres-Douglas theories}
There is another class of SCFTs called the Argyres-Douglas theories which has some different features from the above class of theories \cite{Argyres:1995jj, Argyres:1995xn, Eguchi:1996vu, Eguchi:1996ds}. Their operator spectrum has fractional scaling dimension and there are relevant operators in the spectrum with dimensionful coupling constant. That is, for each relevant operator $u$, there is a coupling constant $m$, such that
their scaling dimensions sum up to $2$, i.e. $[u]+[m]=2$. Such theories can be engineered using the six-dimensional $A_{k-1}$ $(2,0)$ theory on a sphere with an irregular singularity (higher-order pole) \cite{Xie:2012hs}.\footnote{The $A_1$ theory is considered in \cite{Gaiotto:2009hg,Cecotti:2011rv, Bonelli:2011aa}.} There are two kinds of singularity combinations one could use.
\begin{enumerate}
\item[a.] One irregular singularity.
\item[b.] One irregular singularity at the south pole and one regular singularity at the north pole.
\end{enumerate}
The classification of irregular singularities defining the SCFT are studied in \cite{Xie:2012hs}.
Now the coordinates $z$ on the Riemann surface transform nontrivially under the $U(1)_R$ symmetry which gives the fractional scaling dimension. The mass parameter is still encoded in the coefficient of the first-order pole,\footnote{So it is easy to see the appearance of a Higgs branch of the corresponding Argyres-Douglas theory, as studied using other methods in \cite{Argyres:2012fu}.} and the coefficient of the higher-order pole is the dimensionful coupling constant. The operator spectrum needed for calculating the central charges can be found using the following method:
\begin{enumerate}
\item Find the singularity of the Seiberg-Witten curve defining the conformal point, then one can find the scaling dimensions of the coordinates of the curve by
requiring the Seiberg-Witten differential $\lambda=xdz$ to have scaling dimension $1$, since the differential gives the 
mass for the BPS particle.

\item Find the general deformation of the Seiberg-Witten curve, and the scaling dimension of the coefficient of the deformation using the scaling 
dimensions of the coordinates found from 1.
\end{enumerate}
Amazingly, the Argyres-Douglas point and the general deformation can be easily found from the defining data of the irregular singularity much as what happens
in the regular singularity case. Moreover, the 3d mirror (with Lagrangian description) can also be found from the structure of the singularity type, and one could readily find the 
spectrum.

\subsubsection{$\mbox{I}_{k,N}$ theory: $(A_{k-1}, A_{N-1})$ theory}
\begin{figure}
\center
\includegraphics[height=3.5cm]{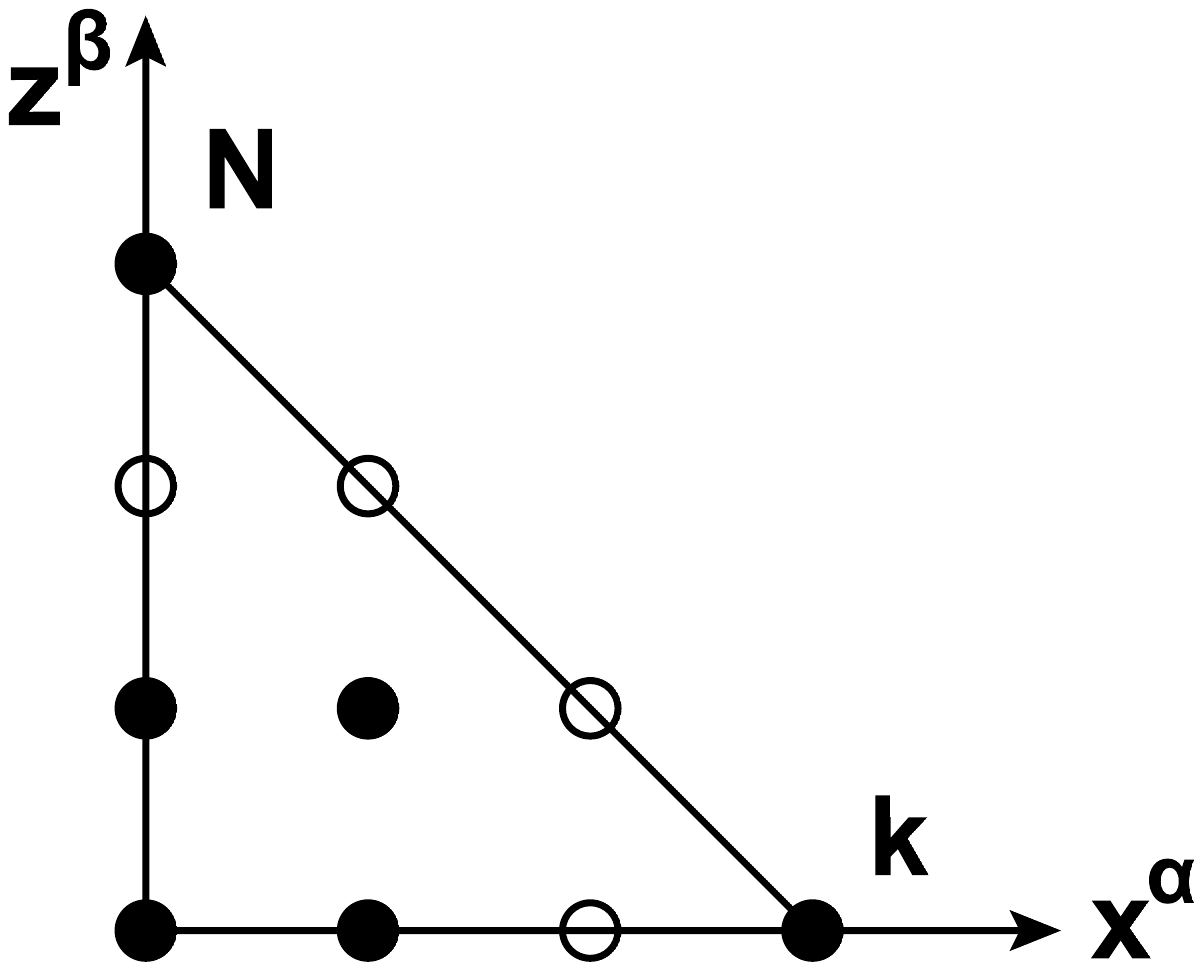}
\caption{The Newton polygon for the $(A_{2}, A_{2})$ theory. The white dots on the lines $\alpha=k-1$ and $\beta=N-1$ are excluded from the spectrum.}
\label{type1}
\end{figure}

This type of irregular singularity can be represented using a Newton polygon shown in figure \ref{type1}: the vertices of the polygon have coordinates 
$(k,0)$ and $(0,N)$. This type of theory is called the $(A_{k-1},A_{N-1})$ theory since its BPS quiver can be written as the direct product of the two Dynkin diagrams \cite{Cecotti:2010fi, Xie:2012dw, Xie:2012jd}. The theory is apparently invariant under exchanging $k$ and $N$.
The Argyres-Douglas point is defined by the following simple equation
\eq{
x^k+z^N=0,
}
and the scaling dimensions of the coordinates are
\eq{
[x]={N\over N+k},\qquad [z]={k\over N+k}.
}
The Seiberg-Witten curve under the general deformation is read from the integer points bounded by the Newton polygon. Each lattice point with coordinates $(\alpha,\beta)$ gives a monomial deformation $x^\alpha z^\beta$ in the Seiberg-Witten curve. We exclude the points on the line $\alpha=k-1$ because the six-dimensional group is $SU(k)$ and the line $\beta=N-1$ using the translational invariance of the $z$ coordinate. 
Once the full Seiberg-Witten curve is written down, one can find the scaling dimension of the operator before each monomial using the above scaling dimension of the coordinates. Let us give a simple example showing how the above idea works in practice.
Consider the $(A_{2}, A_{2})$ theory. We can read off its Seiberg-Witten curve from the Newton polygon in figure \ref{type1}
\eq{ 
x^{3} + x(u_{1}z + u_{2}) + z^{3} + u_{3}z + u_{4} = 0.
} 
The operator spectrum is 
\eq{[u_{1}] = \frac{1}{2}, \qquad [u_{2}] = 1, \qquad [u_{3}] = 1,\qquad [u_{4}] = \frac{3}{2}. } 
We see that this theory has one relevant operator $u_{4}$ and a corresponding coupling constant $u_{1}$. There are also two mass parameters $u_{2}$ and $u_{3}$. 

\begin{figure}
\center
\includegraphics[height=3cm]{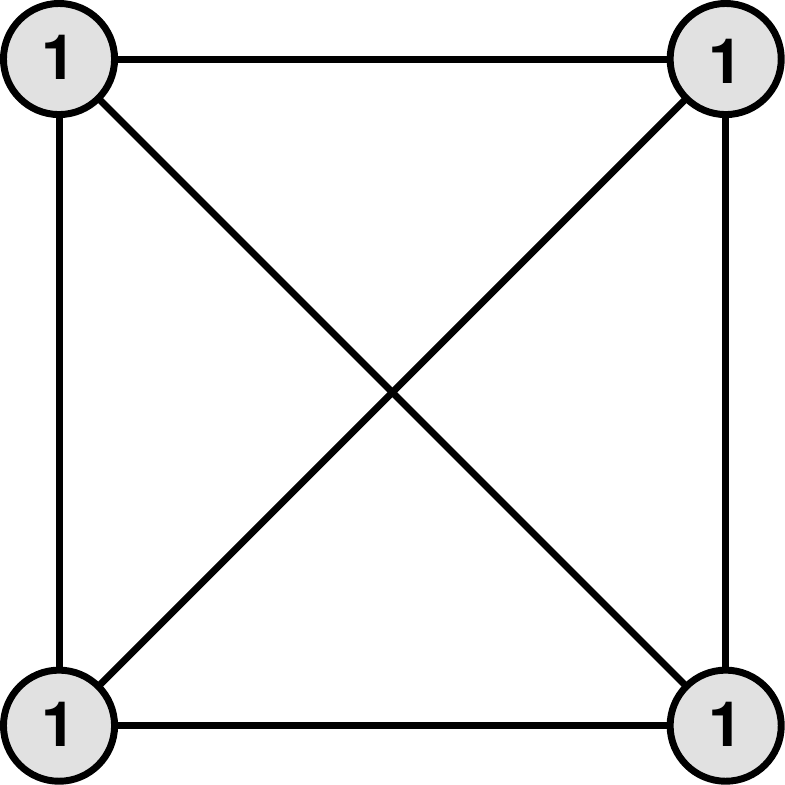}
\caption{The mirror quiver for the $(A_{3}, A_{3})$ theory. The grey nodes denote that the theory corresponds to an irregular singularity.}
\label{I}
\end{figure}

The 3d mirror of the Argyres-Douglas theory is known if $N=kn$: there are $k$ $U(1)$ groups and $n$ arrows between any pair. See figure \ref{I} for the 3d mirror of an Argyres-Douglas theory with $k=4$ and $N=4$. 
The Coulomb branch dimension of the mirror theory is equal to $k-1$.\footnote{Notice that an overall $U(1)$ is decoupled in the 3d mirror theory.} 
The Seiberg-Witten curve can be read from the Newton polygon as
\EQ{&x^{k} + x^{k-2}\left(u_{2,1}z^{2n-1} + \cdots + u_{2,2n}\right) + \cdots + x^{k-l} \left(u_{l,1}z^{ln-1} + \cdots + u_{l,ln}\right) + \cdots \\
&+ (z^{N}+ u_{k,2}z^{N-2} + \cdots + u_{k,N}) = 0.\label{SW1}}
Note that the coefficients of the $x^{k}$ and the $z^{N}$ terms are scaled to be $1$ and the $z^{N-1}$ term is zero by shifting the parameters.
The operator dimensions can be easily found using our general method
\eq{[u_{l,i}] = \frac{i}{n+1}.}
The Higgs branch dimension of the original theory is $k-1$ as seen from the mirror and one can use (\ref{Lagrangian}) to get the following equations for $a$ and $c$:
\eq{2a-c = \frac{1}{4}\sum _{l=2}^k \sum _{i=n+2}^{l n} \left(\frac{2i}{n+1}-1\right), \qquad a - c = -\frac{k-1}{24}.}
We solve the equations to find
\eq{a = \frac{(k-1) \left(2 k^2 n^2+2 k n^2-5 n-5\right)}{24 (n+1)}, \qquad c = \frac{(k-1) \left(k^2 n^2+k n^2-2 n-2\right)}{12 (n+1)}.\label{kn}}
Substituting the above central charges into (\ref{topological}), we can determine $R(B)$. 
\eq{R(B) = \frac{(k-1) k n (k n-1)}{4 (n+1)} = \frac{(k-1) Nk (N-1)}{4 (N+k)}.\label{universal}}
This is manifestly symmetric in $k$ and $N$, which is good since the theory is the same under exchanging $k$ and $N$.
\begin{table}[t]
\begin{center}
\begin{tabular}{c c c}
\hline
Theory & $a$ & $c$ \\ \hline\hline \\
$\text{I}_{2, N_\text{even}}$ & $\dfrac{6 N^2-5 N-10}{24 (N+2)}$ & $\dfrac{3 N^2-2 N-4}{12 (N+2)}$ \\ \\
$\text{I}_{2, N_\text{odd}}$ & $\dfrac{(N-1) (12 N+7)}{48 (N+2)}$ & $\dfrac{(N-1) (3 N+2)}{12 (N+2)}$ \\ \\ \hline \\ 
$\text{I}_{3, N=3n}$ & $\dfrac{8 N^2-5 N-15}{12 (N+3)}$ & $\dfrac{(N+1) (2 N-3)}{3 (N+3)}$ \\ \\
$\text{I}_{3, N={3n-1 \atop 3n-2}}$ & $\dfrac{(N-1) (16 N+11)}{24 (N+3)}$ & $\dfrac{(N-1) (4 N+3)}{6 (N+3)}$
\end{tabular}
\end{center}
\caption{Central charges $(a,c)$ for the type I $(A_{k-1}, A_{N-1})$ theory.}
\label{CCI}
\end{table}
For other theories in this class, there is no 3d mirror. Our proposal of calculating the central charges is based on the assumption that $R(B)$ is valid 
universally for any $k$ and $N$. This is natural since $R(B)$ is derived from the discriminant of the Seiberg-Witten curve which has a universal form. 
Based on the assumption, we now compute the central charges for $N = kn - j$, where $j = 1,\ldots, k-1$. The Seiberg-Witten curve read from the Newton polygon is
\EQ{
&x^{k} + x^{k-2}\left(\cdots\right) + \cdots + x^{k-l} \left(u_{l,\alpha_{l}}z^{ln-\alpha_{l}} + \cdots + u_{l,ln}\right) + \cdots \\
&+ \left(z^{N}+ u_{k,2+j}z^{N-2} + \cdots + u_{k,N+j}\right) = 0.
}
$\alpha_l$ is taken such that $u_{l, \alpha_{l}}$ has a positive scaling dimension. The operator dimensions will be changed to
\eq{[u_{l,i}] = \frac{ik - lj}{k+kn-j}. \label{dim}}
We sum over the operators to find $R(A)$
\eq{R(A) = \sum_{l = 2}^{k} \sum_{i=n+2+\lfloor j(l-1)/k\rfloor}^{ln} \left(\frac{ik - lj}{k+kn-j} - 1\right),\label{RA}}
where $\lfloor x\rfloor$ denotes the largest integer that is smaller than $x$.
Then applying the universal form of $R(B)$ (\ref{universal}) for $N = kn-j$, we can determine the central charges using the general formula (\ref{topological}). For the $k=2$ example, we recover the results in \cite{Xie:2012hs}
\EQ{a &= \frac{12 N^2-10 N+5 j N-10 N \left\lfloor \frac{j}{2}\right\rfloor -3 j^2-12 \left\lfloor \frac{j}{2}\right\rfloor ^2+12 j \left\lfloor \frac{j}{2}\right\rfloor +16 j-32 \left\lfloor \frac{j}{2}\right\rfloor -20}{48 (N+2)}\\
&=\frac{6 N^2-5 N-10}{24 (N+2)} \qquad \text{if } j=0,\\
&=\frac{(N-1) (12 N+7)}{48 (N+2)}\quad \text{if } j=1.
}
The general expression for higher $k$ quickly becomes complicated, but can be computed case by case. We make a list of the central charges for $k=2$ and $k=3$ in table \ref{CCI}. Note that the expression of the central charges as a function of $N$ jumps when $N$ has a common factor with $k$. This is because there are mass parameters for these theories, which have extra degrees of freedom.

\subsubsection{$\mbox{II}_{k,N}$ theory}
\begin{figure}
\center
\includegraphics[height=4cm]{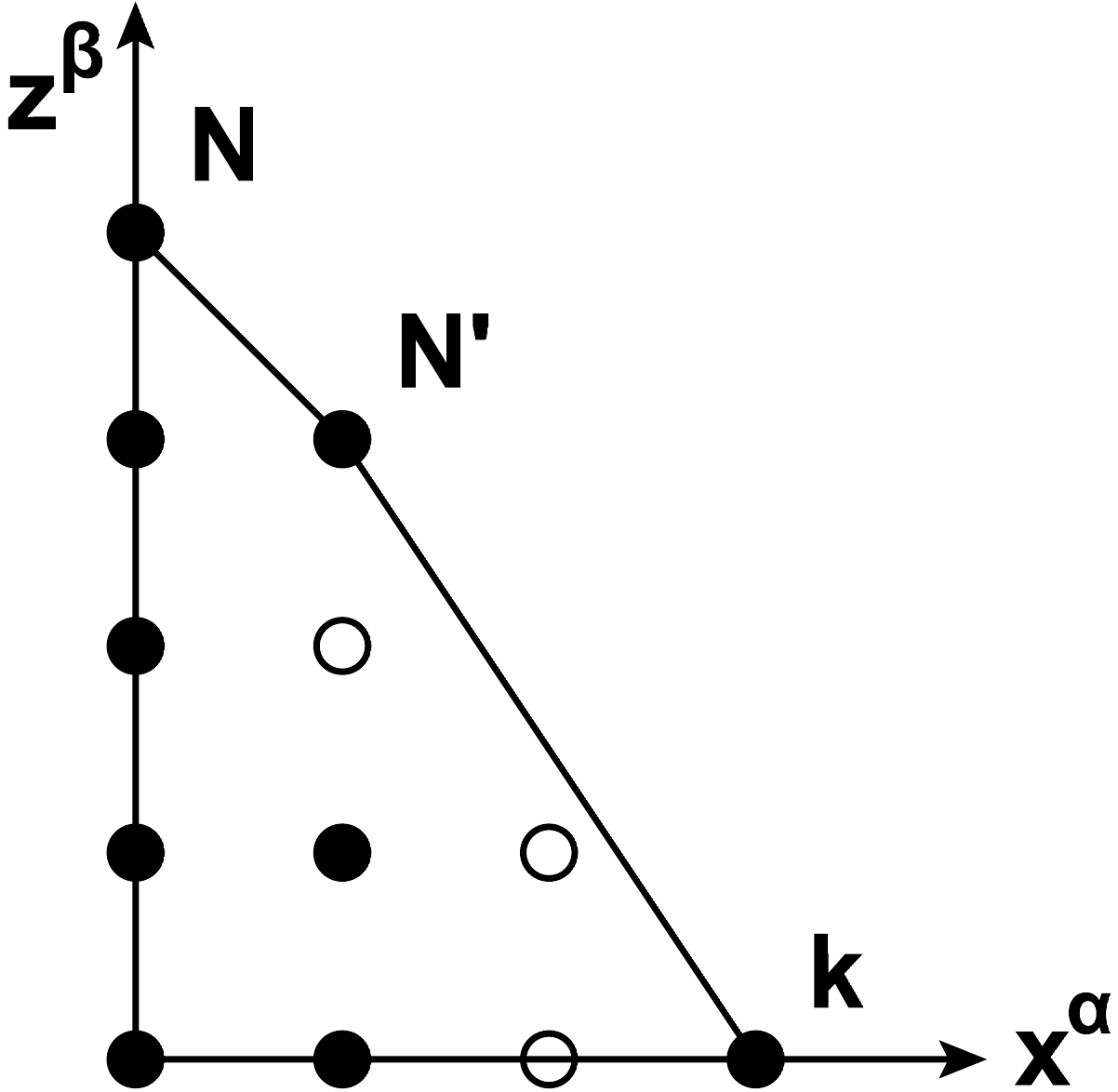}
\caption{The Newton polygon for the $\mbox{II}_{3,4}$ theory.}
\label{type2}
\end{figure}

The Newton polygon for the type II theories is shown in figure \ref{type2}. 
The Argyres-Douglas points are defined by the turning point $(1,N^{'})$ of the Newton polygon 
\eq{
x^k+xz^{N^{'}}=0.
}
The top point of the Newton polygon with 
coordinates $(0,N)$ is determined by $N^{'}$ as 
\eq{N=N^{'}+\left\lfloor{N^{'}\over k-1}\right\rfloor.\label{NN'}}
The slope of the second segment of the Newton polygon is equal to the integer part of the slope of the first segment. 
The Seiberg-Witten curve is 
\EQ{&x^{k} + x^{k-2}(\cdots) + \cdots + x\left(z^{N^{'}}+u_{k-1,2}z^{N^{'}-2} + \cdots + u_{k-1,N^{'}}\right)+ \\
&+ \left(u_{k,0}z^{N} + u_{k,1}z^{N-1} + \cdots + u_{k,N}\right) = 0.}
When $N^{'}=(k-1)n$ is an integer, this theory is the same as the $\mbox{I}_{k,N}$ theory. 

Assuming that $R(B)$ is a universal function of $N^{'}$, we can compute the central charges for any $N^{'}$. We list the central charges for $k=2$ and $k=3$ in table \ref{CCII}. Notice that $N$ can only take certain values due to the relation between $N^{'}$ and $N$ (\ref{NN'}). One may easily check that the central charges for $N = kn$ coincide with table \ref{CCI}.
\begin{table}
\begin{center}
 \begin{tabular}{c c c}
 \hline
 Theory & $a$ & $c$ \\ \hline\hline \\
 $\text{II}_{3,N=3n}$ & $\dfrac{12 {N^{'}}^{2}-5 N^{'}-10}{12 \left(N^{'}+2\right)}$ & $\dfrac{\left(N^{'}-1\right) \left(3 N^{'}+2\right)}{3 \left(N^{'}+2\right)}$ \\ \\
 $\text{II}_{3,N=3n-2}$ & $\dfrac{16 {N^{'}}^{2}-5 N^{'}-9}{16 \left(N^{'}+2\right)}$ & $\dfrac{4 {N^{'}}^{2}-N^{'}-2}{4 \left(N^{'}+2\right)}$
 \end{tabular}
 \end{center}
 \caption{Central charges $(a,c)$ for the type $\text{II}_{3,N}$ theory. When $N=3n$, we have $N^{'}=2n={2\over3}N$, and the $\text{II}_{3,N}$ theory is isomorphic to the $\text{I}_{3,N}$ theory. One can check that their central charges are the same by comparing with table \ref{CCI}.}
 \label{CCII}
\end{table}
As we discuss later, this class of theories is actually isomorphic to one class of type IV theories defined by adding an extra regular singularity. 

\subsubsection{Type III theory: degenerating case}
If the pole is integer for the type I singularity, the eigenvalues of the coefficient of the irregular singularity pole can be degenerate and 
this defines a new class of theories. The defining data for the irregular singularity involves a sequence of Young tableaux $Y_{n+2} \subseteq Y_{n+1} \subseteq \cdots \subseteq Y_{1}$,\footnote{This theory is a degenerating case of the $\mbox{I}_{k,kn}$ Argyres-Douglas theory.} and each tableau $Y_{j}$ is obtained from further partitioning the previous tableau $Y_{j+1}$. These Young tableaux specify the eigenvalue degeneracy of the pole matrices of the Higgs field in the Hitchin system. 

The Seiberg-Witten curve for the type III theory is the same as the type I theory, but not all the deformations are independent. The independent operators are counted as follows: to each tableau $Y_{j}$ we associate a vector $p^{(j)}$ whose $i$-th component is 
\eq{p^{(j)}_{i} = i - s^{(j)}_{i},}
where $s^{(j)}_{i}$ is the height of the $i$-th box of $Y_{j}$. The Seiberg-Witten curve has the familiar form
\eq{
x^k+\sum_{i=2}^k\Phi_i(z)x^{k-i}=0,
}
where $\Phi_i(z)$ is a polynomial in $z$. We want to determine a cutoff number $d_i$ for each $\Phi_i(z)$ such that we only count the operators 
from the coefficient of $x^{k-i}z^j$ with $j\leq d_i$. $d_{i}$ is determined by summing $p^{(j)}_{i}$ over all the Young tableaux as
\eq{d_{i} = \sum^{n+2}_{j=1} p^{(j)}_{i} -2i + 1. \label{cutoff}}
This number may be negative. Here we only consider the case where $d_i\geq 0$ for simplicity. Other cases can be analyzed in a similar way.

The rule for finding the 3d mirror for an irregular puncture is a little more complicated than for a regular puncture, but can be done step by step like atomic fission.
First to $Y_{n+2}$ we associate a quiver with nodes each labeled by the height of each column in $Y_{n+2}$. The nodes are connected to one another by $n$ lines. 
Then we split each node into nodes of lower rank according to the form of $Y_{n+1}$: if one column with height $m$ in $Y_{n+2}$ is further partitioned as several columns with height $[m_1,\ldots, m_s]$, then
one quiver node with rank $m$ is decomposed into $s$ nodes with rank $(m_1, \ldots, m_s)$. The newly created nodes are connected to one another by $n-1$ lines. We continue the procedure until $Y_{2}$. In the last step, instead of splitting the nodes further, we attach a quiver tail according to the change from $Y_{2}$ to $Y_{1}$: if a column with height $h$ in $Y_2$ is decomposed into several columns $[h_1,\ldots,h_r]$, we then attach a quiver tail as we do for a regular puncture with the Young tableau $[h_1,\ldots,h_r]$.

\begin{figure}
\center
\includegraphics[width=12cm]{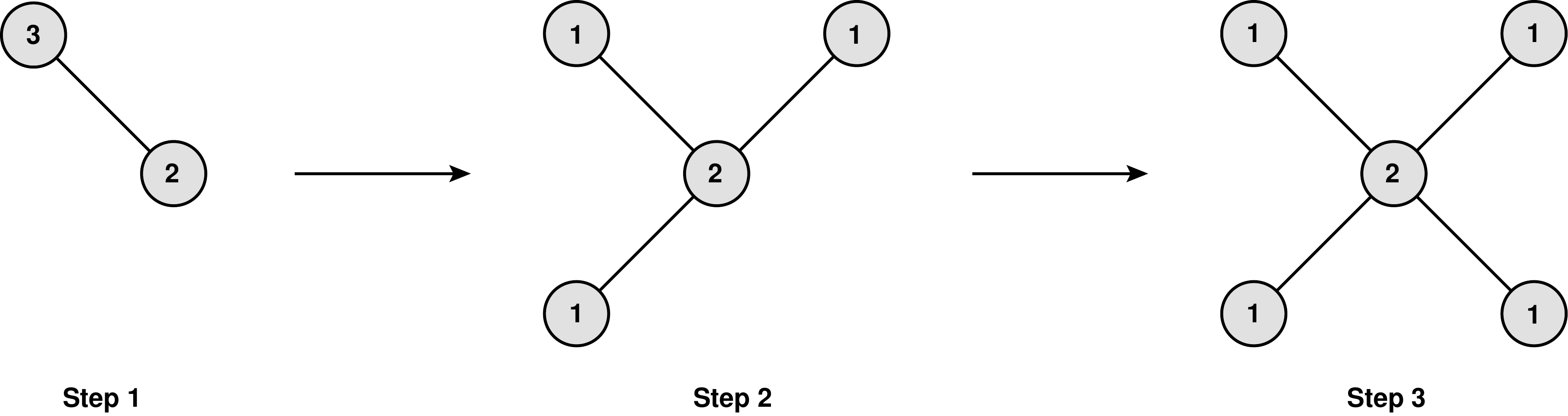}
\caption{The mirror pair of the $SU(2)$ superconformal QCD from an irregular puncture.}
\label{irreg}
\end{figure}

For example, let us consider the sequence $Y_{3} = \tiny\yng(1,2,2), \quad Y_{2}= \tiny\yng(1,4), \quad Y_{1}=\tiny\yng(5)$. The columns of $Y_{3}$ have heights 3 and 2 so give a $3-2$ quiver as shown in figure \ref{irreg}. To obtain $Y_{2}$, we split the column of height 3 into three columns of height 1. Finally, we split the column of height 2 into two columns of height 1. Because this is the last step, we get a $\boxed{2}-1$ quiver tail which we attach to the single $SU(2)$ node by gauging the flavor symmetry. Note that the resulting mirror pair coincides with that of the $SU(2)$ superconformal QCD with $N_{f} = 4$ quarks. This gives us hints that theories defined with irregular singularities can also be used to define theories previously defined with regular singularities. This is actually true: every theory defined using the sphere with regular punctures has an irregular realization.
Having described how to find the operator spectrum and the 3d mirror, we can then use the formula (\ref{Lagrangian}) to find their central charges.

\paragraph{Example 3:} Let us consider a simple example where every Young tableau has the form $[k,k]$. The number $d_i$ from the formula (\ref{cutoff}) is 
\eq{
d_i=(n+2)\left\lfloor\frac{i}{2}\right\rfloor-2i+1.
}
It determines the subset of Coulomb branch operators we sum over 
\eq{
2a - c = \frac14 \sum^{k}_{l=2}\sum^{ln}_{i=ln-d_{l}+1}\left(\frac{2i}{n+1} - 1 \right). 
}
The 3d mirror has only two nodes with gauge group $U(k)$, and its Coulomb branch dimension is $2k-1$. Using the formula (\ref{Lagrangian}), we find the central charges
\EQ{
a&= \frac{6 k^3 n^2-8 k^3 n-8 k^3+6 k^2 n^2+4 k n+4 k-n-1}{24 (n+1)}, \\ 
c&= \frac{(k+1) \left(3 k^2 n^2-4 k^2 n-4 k^2+4 k n+4 k-n-1\right)}{12 (n+1)}.
}
\subsubsection{Type IV theory}\label{typeIV}
\begin{figure}
\center
\includegraphics[height=4cm]{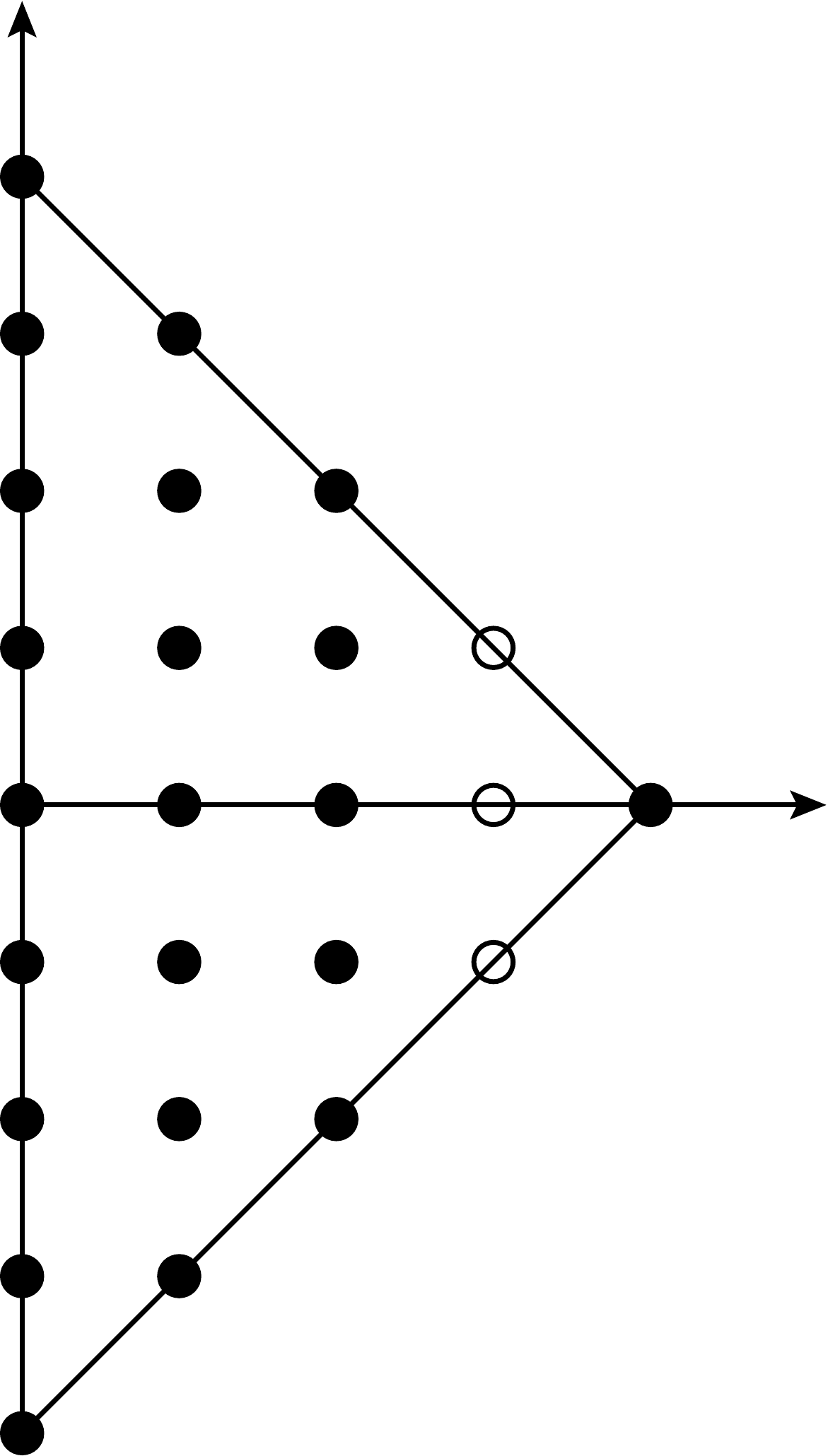} \qquad \qquad \qquad 
\includegraphics[height=4cm]{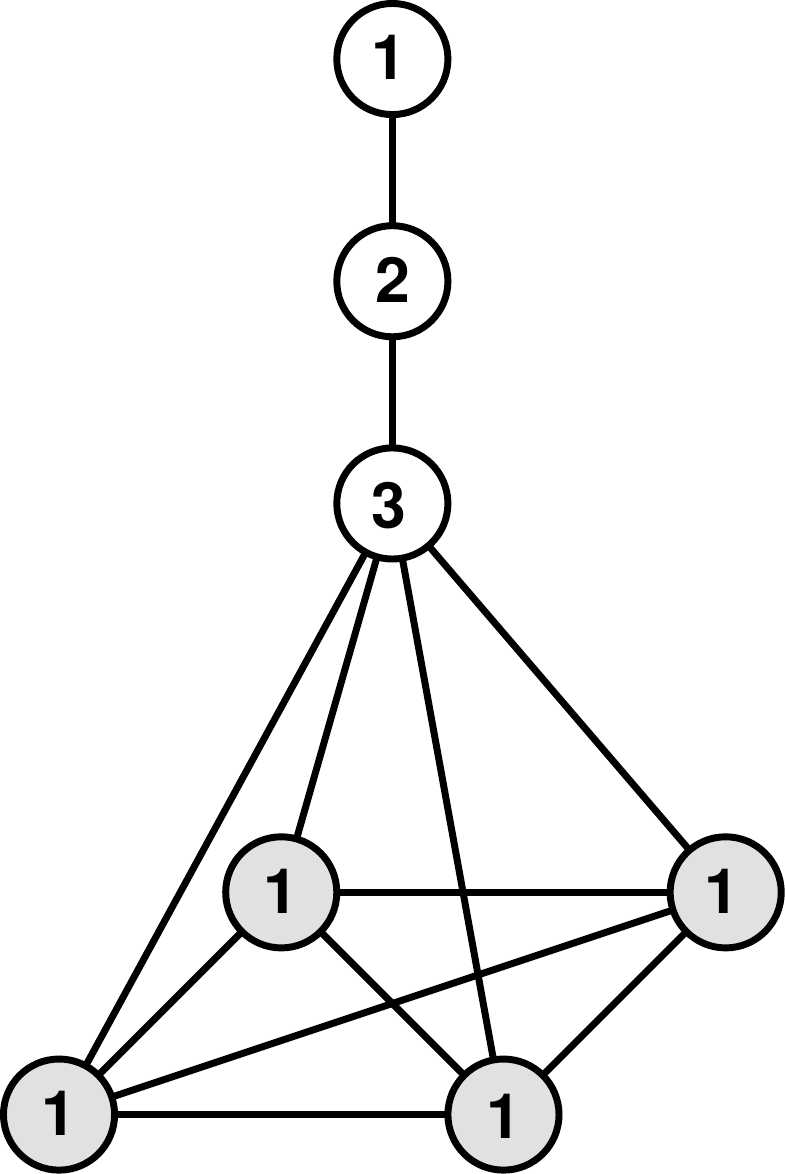}
\caption{The Newton polygon and the mirror pair of the $(\text{I}_{4, 4}, F)$ theory. The white nodes come from the regular puncture, while the grey nodes come from the irregular puncture.}
\label{IF}
\end{figure}

One can add a regular singularity to the above three types of irregular singularities, and define another class of Argyres-Douglas theories with non-abelian flavor symmetry.
If the original irregular singularity has a 3d mirror quiver description, then so does the new theory. We associate a quiver tail to the regular singularity, and 
then spray the $U(k)$ flavor group into subgroups determined by the pattern of the Young tableau $Y_2$ of the irregular singularity, and then glue them together.
The operator spectrum can also be found using the Newton polygon: an extra line with slope 1 is drawn below the positive $x$ axis. See figure \ref{IF}. 
The method of finding the central charges $a$ and $c$ are quite similar. We show some examples below. 

\paragraph{$(\mbox{I}_{k,N},F)$ theory:}
Let us add a full puncture $Y_{0} = [1, \ldots, 1]$ to the irregular singularity defining the $\text{I}_{k,N}$ theory. We call it the $(\text{I}_{k, N}, F)$ theory. The Seiberg-Witten curve at the Argyres-Douglas point is the same as the $\text{I}_{k,N}$ theory, and the analysis of the spectrum is very similar. For $N=kn \ (j=0)$, the Seiberg-Witten curve is
\EQ{
&x^{k} + \cdots + x^{k-l} \left(u_{l, 1}z^{ln-1} + \cdots + \frac{u_{l, l(n+1)}}{z^{l}}\right) + \cdots \\
&+ \left(z^{N}+ u_{k, 1}z^{N-1} + \cdots + \frac{u_{k, N+k}}{z^k}\right)= 0.
\label{generic}
} 
The crucial difference is that the deformation corresponding to the points on the line $\beta=N-1$ should be turned on because the the extra regular puncture breaks the translational invariance of the $z$ coordinate.
See figure \ref{IF} for the Newton polygon and the mirror pair of the $(\text{I}_{4,4}, F)$ theory.
The Coulomb branch dimension of the mirror theory is $\frac{(k-1)(k+2)}{2}$. 
We learn from (\ref{Lagrangian}) that
\eq{2a-c = \frac{1}{4}\sum _{l=2}^k \sum _{i=n+2}^{l (n+1)-1} \left(\frac{2i}{n+1}-1\right), \qquad a - c = -\frac{(k-1)(k+2)}{48}.}
Note that we do not include $u_{l,l(n+1)}$ because they are mass terms with scaling dimensions $l$. We solve the equations to find the central charges and via (\ref{topological}) the universal form of $R(B)$
\EQ{
a &= \frac{1}{48} (k-1) \left(4 k^2 n+4 k^2+4 k n-k-10\right), \\
c &= \frac{1}{12} (k-1) \left(k^2 n+k^2+k n-2\right), \\
R(B) &= \frac{1}{4} (k-1) k (N+k-1).\label{IVfull}}

\begin{figure}
\center
\includegraphics[height=3cm]{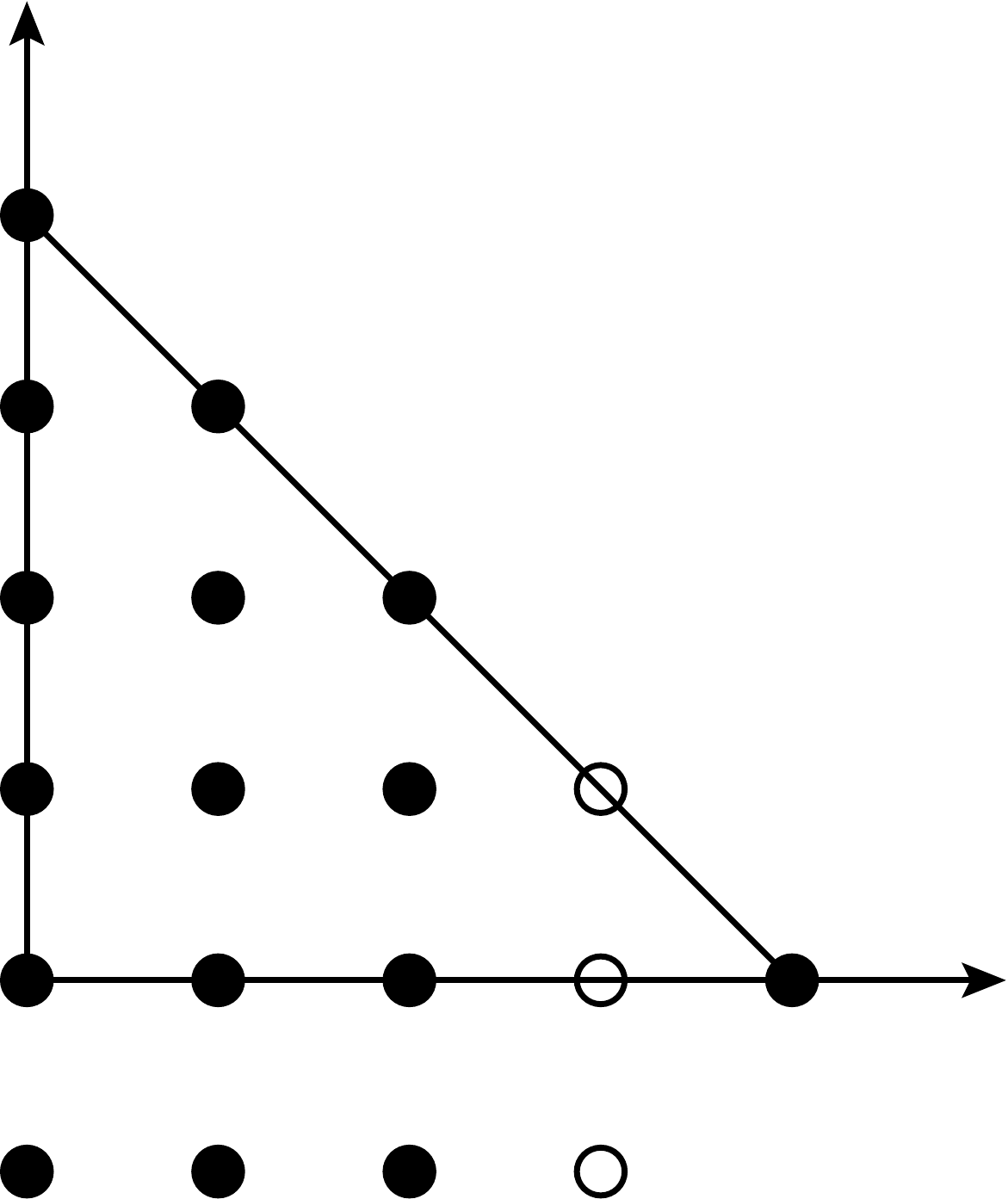} \qquad \qquad \qquad 
\includegraphics[height=3cm]{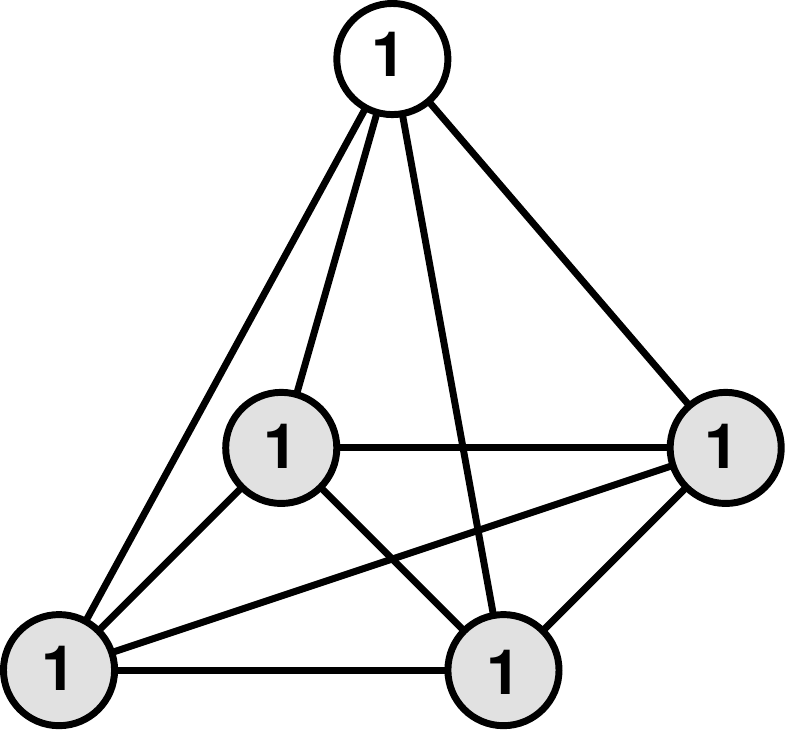}
\caption{The Newton polygon and the mirror pair of the $(\text{I}_{4, 4}, S)$ theory.}
\label{IS}
\end{figure}

\begin{figure}
\center
\quad\includegraphics[height=3.5cm]{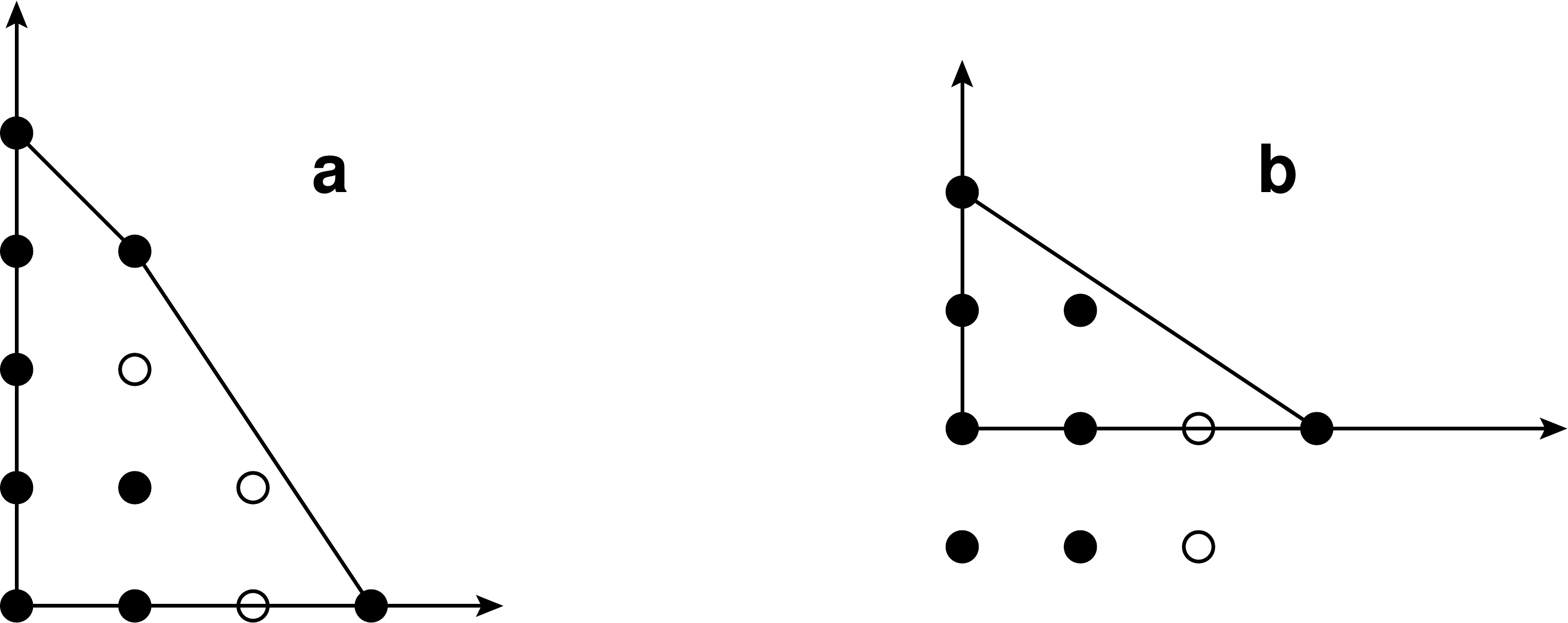}
\caption{The Newton polygons for the a) $\mbox{II}_{3,4}$ theory, b) $(\mbox{I}_{3,2},S)$ theory.}
\label{type24}
\end{figure}

\paragraph{$(\text{I}_{k,N},S)$ theory:} Let us add a simple puncture with the Young tableau $Y_{0} = [k-1,1]$ to the type I irregular singularity. The Newton polygon and the mirror pair for the $(\text{I}_{4,4},S)$ theory is shown in figure \ref{IS}. 
This theory is isomorphic to the $\mbox{II}_{N+1, k+\left\lfloor{k\over N}\right\rfloor}$ theory as can be seen by looking at the spectrum. 
See figure \ref{type24}. 
The Seiberg-Witten curve still has the same form as (\ref{generic}), but only the coefficients of the ${1\over z}$ term are counted as independent 
Coulomb branch operators. For $N=kn \ (j=0)$, the Coulomb branch dimension of the mirror theory is $k$, and we have
\eq{2a-c = \frac{1}{4}\sum _{l=2}^k \sum _{i=n+2}^{ln+1} \left(\frac{2i}{n+1}-1\right), \qquad a - c = -\frac{k}{24}.}
The central charges and the universal form of $R(B)$ are
\EQ{
a&=\frac{k \left(2 k^2 n^2+6 k n-2 n^2-5 n+1\right)}{24 (n+1)}, \\
c&=\frac{k \left(k^2 n^2+3 k n-n^2-2 n+1\right)}{12 (n+1)},\\
R(B)&= \frac{k (N+1)(kN-N+k)}{4(N+k)}.}

Applying $R(B)$ to other cases, we find the central charges for theories with $N=kn-j$ as before. 
We make a list for $k=2$ and $k=3$ in table \ref{CCIV}. Note that when $k=2$, a simple puncture is also a full puncture, so their central charges agree. We recover the central charges for the $(A_{1}, D_{N+2})$ theories \cite{Xie:2012hs}.\footnote{The label $(A_{1}, D_{N+2})$ means that the BPS quiver of the 
theory has the $D_{N+2}$ shape. This theory is labeled as $(\mbox{I}_{2,N},F)$ in this paper.}

\begin{table}
\begin{center}
 \begin{tabular}{c c c}
 \hline
 Theory & $a$ & $c$ \\ \hline\hline \\
 $(\text{I}_{2,N_\text{even}}, F)$ & $\dfrac{1}{12} (3 N+1)$ & $\dfrac{1}{12} (3 N+2)$ \\ \\
 $(\text{I}_{2,N_\text{odd}}, F)$ & $\dfrac{(N+1) (4 N+7)}{16 (N+2)}$ & $\dfrac{N+1}{4}$ \\ \\ \hline \\ 
 $(\text{I}_{3,N=3n}, F)$ & $\dfrac{1}{24} (16 N+23)$ & $\dfrac{1}{6} (4 N+7)$ \\ \\
 $\left(\text{I}_{3,N={3n-1 \atop 3n-2}}, F\right)$ & $\dfrac{(N+2) (4 N+11)}{6 (N+3)}$ & $\dfrac{2 (N+2)}{3}$ \\ \\ \hline \\ 
 $(\text{I}_{3,N=3n}, S)$ & $\dfrac{16 N^2+39 N+9}{24 (N+3)}$ & $\dfrac{8 N^2+21 N+9}{12 (N+3)}$ \\ \\
 $\left(\text{I}_{3,N={3n-1 \atop 3n-2}}, S\right)$ & $\dfrac{(N+1) (4 N+7)}{6 (N+3)}$ & $\dfrac{(N+1) (8 N+15)}{12 (N+3)}$
 \end{tabular}
 \end{center}
 \caption{Central charges $(a,c)$ for the $\text{I}_{k, N}$ theory with a full/simple regular puncture.}
 \label{CCIV}
\end{table}

\paragraph{$(\mbox{II}_{k,N},S)$ and $(\mbox{II}_{k,N},F)$ theory:} As with the type I irregular singularity, we can add a simple or a full regular singularity to the type II irregular singularity. 
\begin{figure}
\center
\includegraphics[scale=0.3]{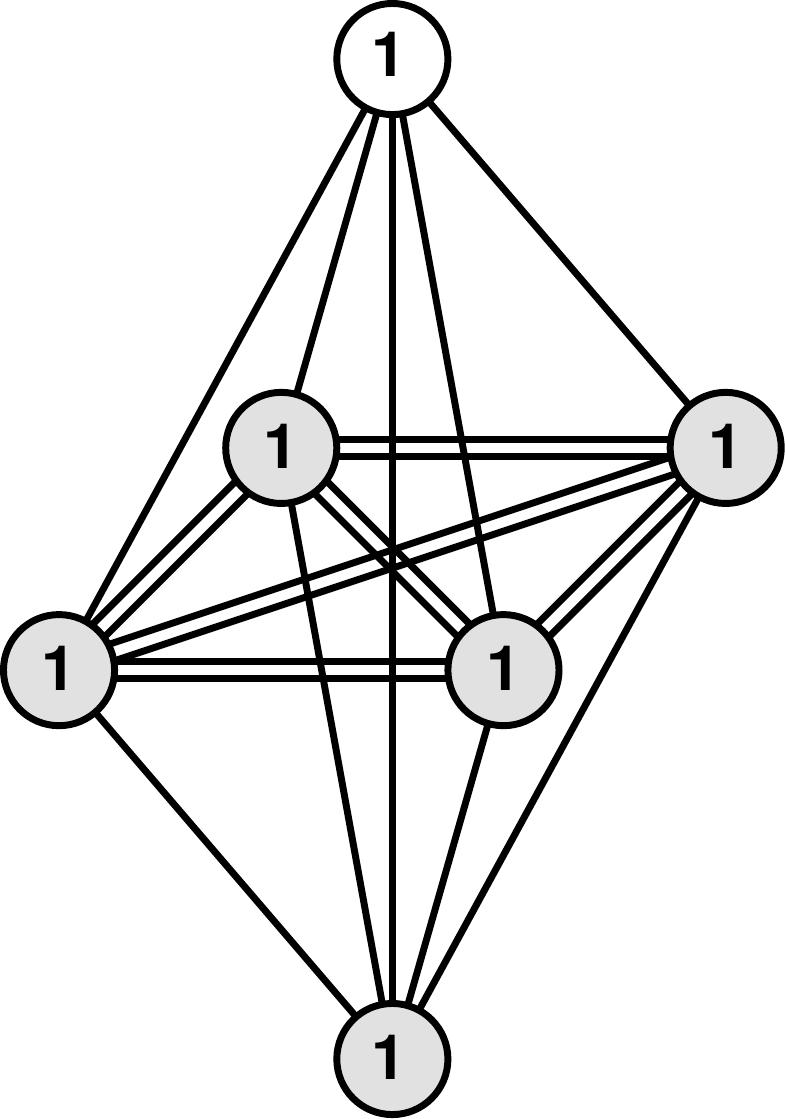} \qquad \qquad \qquad \includegraphics[scale=0.3]{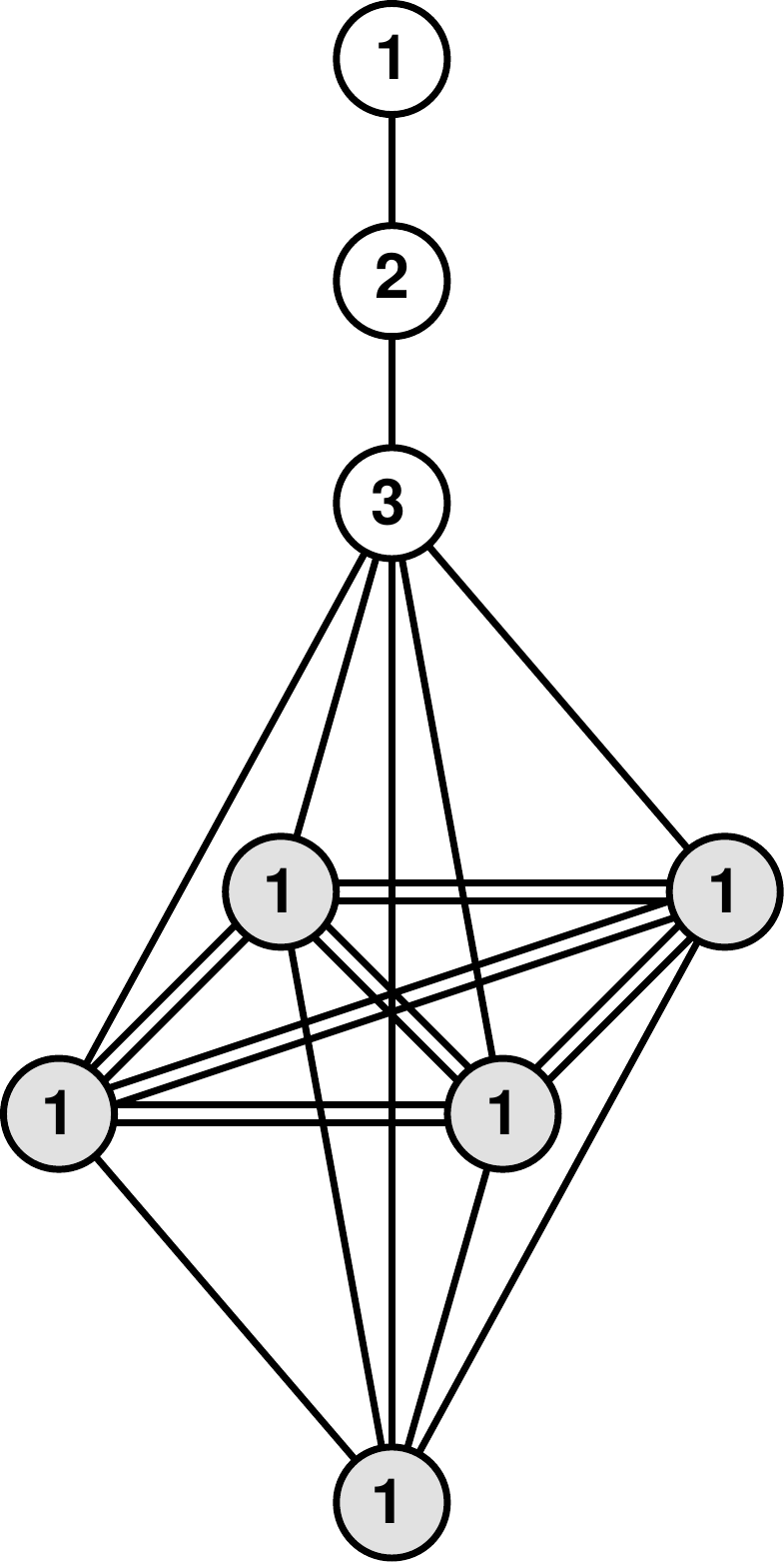}
\caption{Mirror pairs of the $(\text{II}_{9, 4}, S)$ theory and the $(\text{II}_{9, 4}, F)$ theory.}
\label{IImirror}
\end{figure}
The central charges can be calculated in exactly the same way as before. See table \ref{CCIVI}.
\begin{table}
\begin{center}
 \begin{tabular}{c c c}
 \hline
 Theory & $a$ & $c$ \\ \hline\hline \\
 $(\text{II}_{2,N_\text{even}}, F)$ & $\dfrac{1}{12} \left(6 N^{'}+1\right)$ & $\dfrac{1}{6} \left(3 N^{'}+1\right)$ \\ \\ \hline \\ 
 $(\text{II}_{3,N=3n}, F)$ & $\dfrac{1}{24} \left(24 N^{'}+23\right)$ & $\dfrac{1}{6} \left(6 N^{'}+7\right)$ \\ \\
 $(\text{II}_{3,N=3n-2}, F)$ & $\dfrac{16 {N^{'}}^2+49 N^{'}+35}{16 \left(N^{'}+2\right)}$ & $\dfrac{1}{4} \left(4 N^{'}+5\right)$ \\ \\ \hline \\
 $(\text{II}_{3,N=3n}, S)$ & $\dfrac{8 {N^{'}}^2+13 N^{'}+2}{8 \left(N^{'}+2\right)}$ & $\dfrac{4 {N^{'}}^2+7 N^{'}+2}{4 \left(N^{'}+2\right)}$ \\ \\
$(\text{II}_{3,N=3n-2}, S)$ & $\dfrac{48 {N^{'}}^2+83 N^{'}+25}{48 \left(N^{'}+2\right)}$ & $\dfrac{\left(2 N^{'}+1\right) \left(3 N^{'}+4\right)}{6 \left(N^{'}+2\right)}$
 \end{tabular}
 \end{center}
 \caption{Central charges $(a,c)$ for the type $\text{II}_{k,N}$ theory with a full/simple regular puncture.}
 \label{CCIVI}
\end{table}

\subsection{General features of the central charges}
\begin{figure}
\begin{center}
\includegraphics[width=8.5cm]{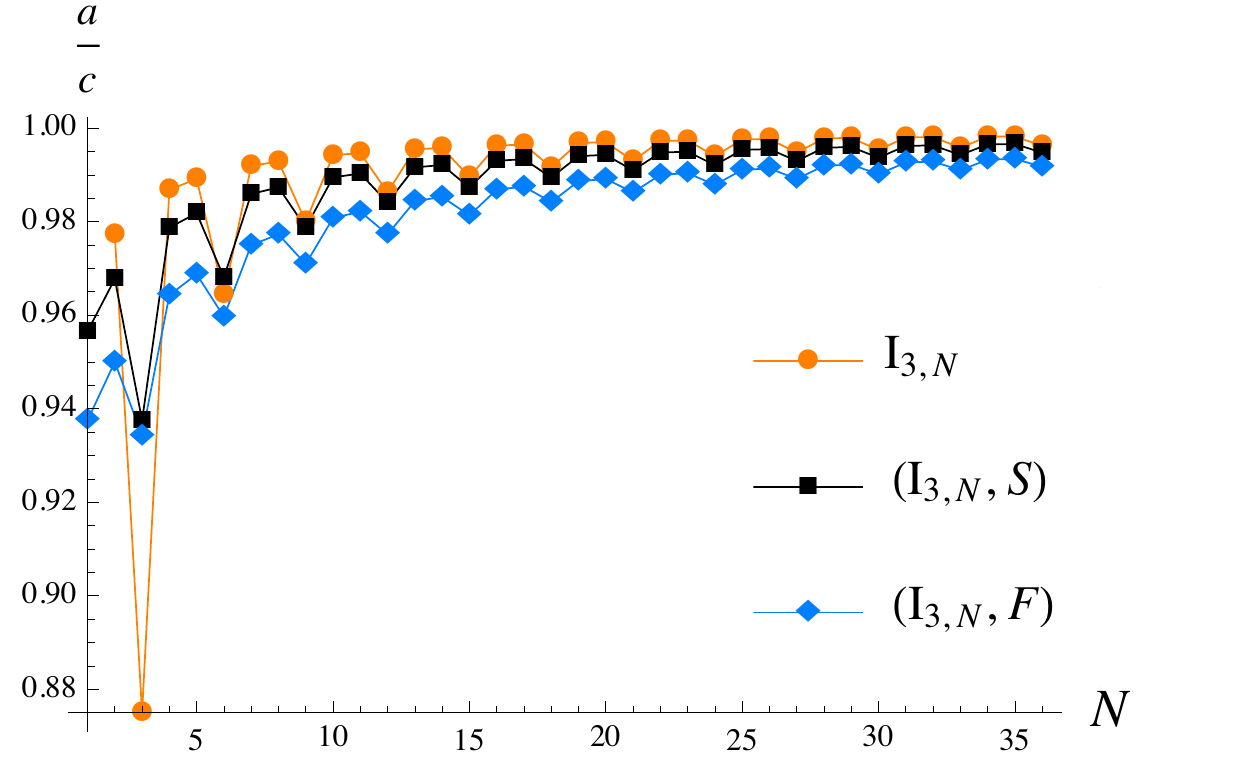}\!\!\!\!\!\!\!\!\!\!\includegraphics[width=8.25cm]{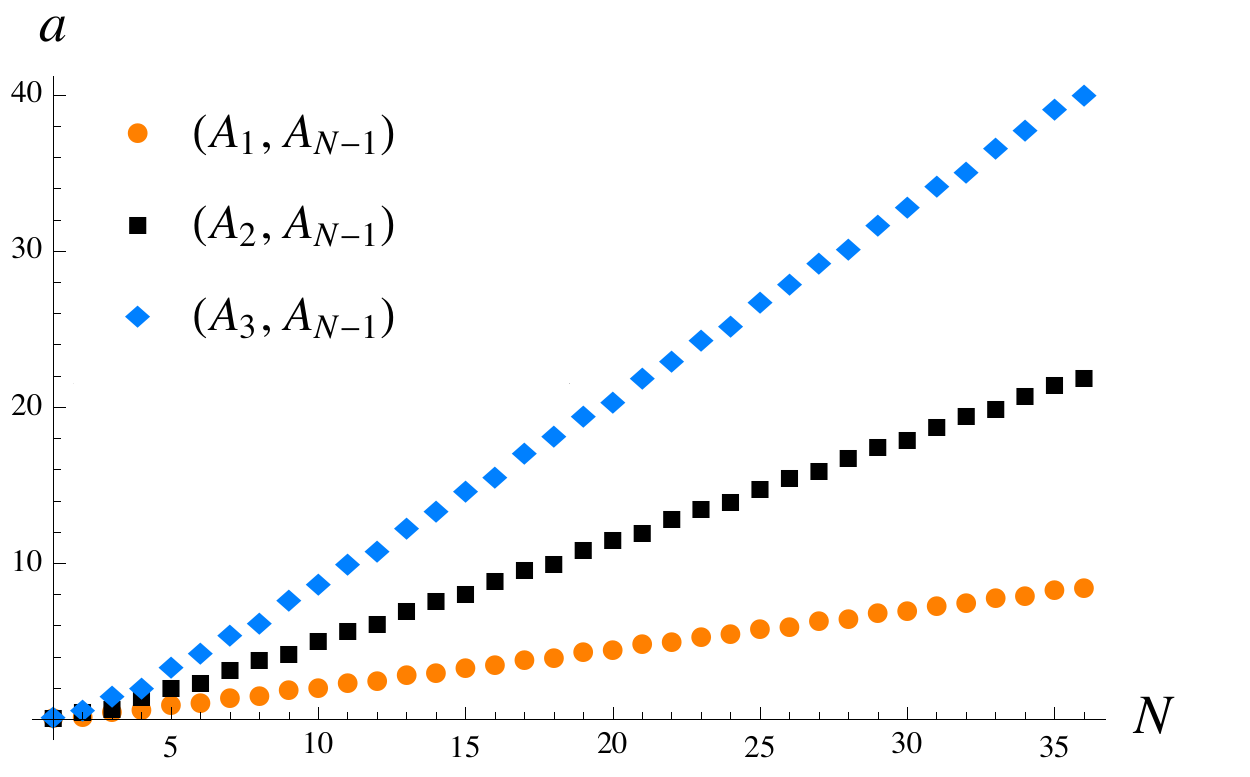}
\caption{a) The ratio $a/c$ asymptotes to 1 at large $N$. b) Central charge $a$ for type $\text{I}_{k,N}$ theories.}\label{ratio}
\end{center}
\end{figure}

The ratio of central charges $a/c$ is of interest and it is bounded for any $\mathcal{N}=2$ theory
by \cite{Shapere:2008zf}
\eq{
\frac12 \le \frac{a}c \le \frac54.}
This bound is obtained based on a positive energy assumption in \cite{Hofman:2008ar}, and it is derived using topological gauge theories in \cite{Shapere:2008zf}.
The lower bound is attained by free hypermultiplets and the upper bound by free vectormultiplets. The central charge ratios of all our theories fall inside this bound, which strongly confirms that our methods are correct. Moreover, they all approach 1 at large $N$, which suggests that there should be a nice supergravity dual. See figure \ref{ratio}a.

We can also plot the central charge $a$ as a function of $N$ for each type of theory and see that it indeed decreases monotonically as $N$ decreases. See figure \ref{ratio}b. Although the central charge $c$ is not guaranteed to be monotonic in general, we find that it has a similar behavior as the central charge $a$ for our class of theories because the ratio $a/c$ rapidly approaches 1. In the next subsection, we are going to study the RG flow between these theories and check that it satisfies the $a$-theorem.

\subsection{RG flow of general Argyres-Douglas theories}
For the Argyres-Douglas theory, there are relevant operators in the spectrum. One can turn on the relevant deformation and flow to a new fixed point in the infrared, i.e. one can turn on the deformation \cite{Argyres:1995xn}
\eq{
\int d^4 \theta~{\langle v\rangle\over \mu^{\sigma}} U,
}
where $\langle v\rangle$ is the expectation value of certain operator with dimension $1$. The coupling constant is identified as $m={\langle v\rangle\over \mu^{\sigma}}$ with scaling dimension 
\eq{
[m]=1-\sigma,
}
and the operator $U$ has dimension $[U]=1+\sigma$. $U$ is an irrelevant operator when $[U]>2$, marginal when $[U]=2$, and relevant when $[U]<2$. 

We want to find the nearest IR fixed point with the minimal decrease in the central charge $a$ by turning on a relevant operator. This can be easily done by looking at the Seiberg-Witten curve under the corresponding deformation. Let us look at the Argyres-Douglas point of the $(A_{k-1}, A_{N-1})$ theory:
\eq{
x^k+z^N=0.
}
If the coupling constant deformation for the relevant operator corresponds to turning on the term $z^{N-2}$, then the Seiberg-Witten curve is 
\eq{
x^k+z^N+{\langle v \rangle\over \mu^{\sigma}}z^{N-2}=0, \label{deformation}
}
and $\sigma={N-k\over N+k}$. In the deep IR as $\mu\rightarrow 0$, one needs to take a scaling limit 
\eq{x\rightarrow \left(\langle v \rangle \mu^{\sigma_1}\right)^{1\over k}x^{'}, \qquad z \rightarrow \left(\mu^{\sigma_2}\right)^{\frac{1}{N-2}}z^{'}.}
We require that the $\mu$ factor for $x^k$ and $z^{N-2}$ are the same, and that the sum of the new coordinates have dimension 1. Therefore we have the equations
\EQ{
&\sigma_1=\sigma_2-\sigma=\sigma_2-{N-k\over N+k}, \\
& (1+\sigma_1){1\over k}+\sigma_2{1\over N-2}=0.
}
Solving the equations, we get 
\eq{
\sigma_1=\frac{k^2-2 k N+2 k-N^2+2 N}{(N+k-2) (N+k)},\qquad \sigma_2=-\frac{2 k (N-2)}{(N+k-2) (N+k)}.
\label{solution}
}
The second term in (\ref{deformation}) can be ignored in the limit $\mu\rightarrow 0$, and we get the new Argyres-Douglas point 
\eq{
x^{'k}+z^{'N-2}=0.
}
It is easy to check using the solution (\ref{solution}) that $x^{'}$ and $z^{'}$ indeed have scaling dimensions $\frac{N-2}{N+k-2}$ and ${k\over N+k-2}$, respectively; so the new IR fixed point is the $(A_{k-1}, A_{N-3})$ theory. For some other cases, the least relevant deformation corresponds to turning on the term $xz^{N-2}$. Using exactly the same logic as the above analysis, we get the new fixed point
\eq{
x^k+xz^{N-2}=0,
}
which is a type II theory. 
 
Let us use the above method to find the least RG flow for other Argyres-Douglas theories. For the type II theory whose Argyres-Douglas point is 
\eq{
x^k+xz^{N^{'}}=0,
}
the least relevant deformation corresponds to the monomial $z^{N^{'}+\lfloor{\frac{N^{'}}{k-1}\rfloor}}$. The IR fixed point would be a type I theory
\eq{
x^k+z^{N^{'}+\lfloor{\frac{N^{'}}{k-1}\rfloor}}=0.
}
When flowing between type I theories, $N$ decreases by at least 2 because the deformation corresponding to the term $z^{N-1}$ is excluded. Similarly, when we flow from one type II theory to another, $N^{'}$ decreases by at least 2.

\begin{figure}
\center
\includegraphics[scale=0.3]{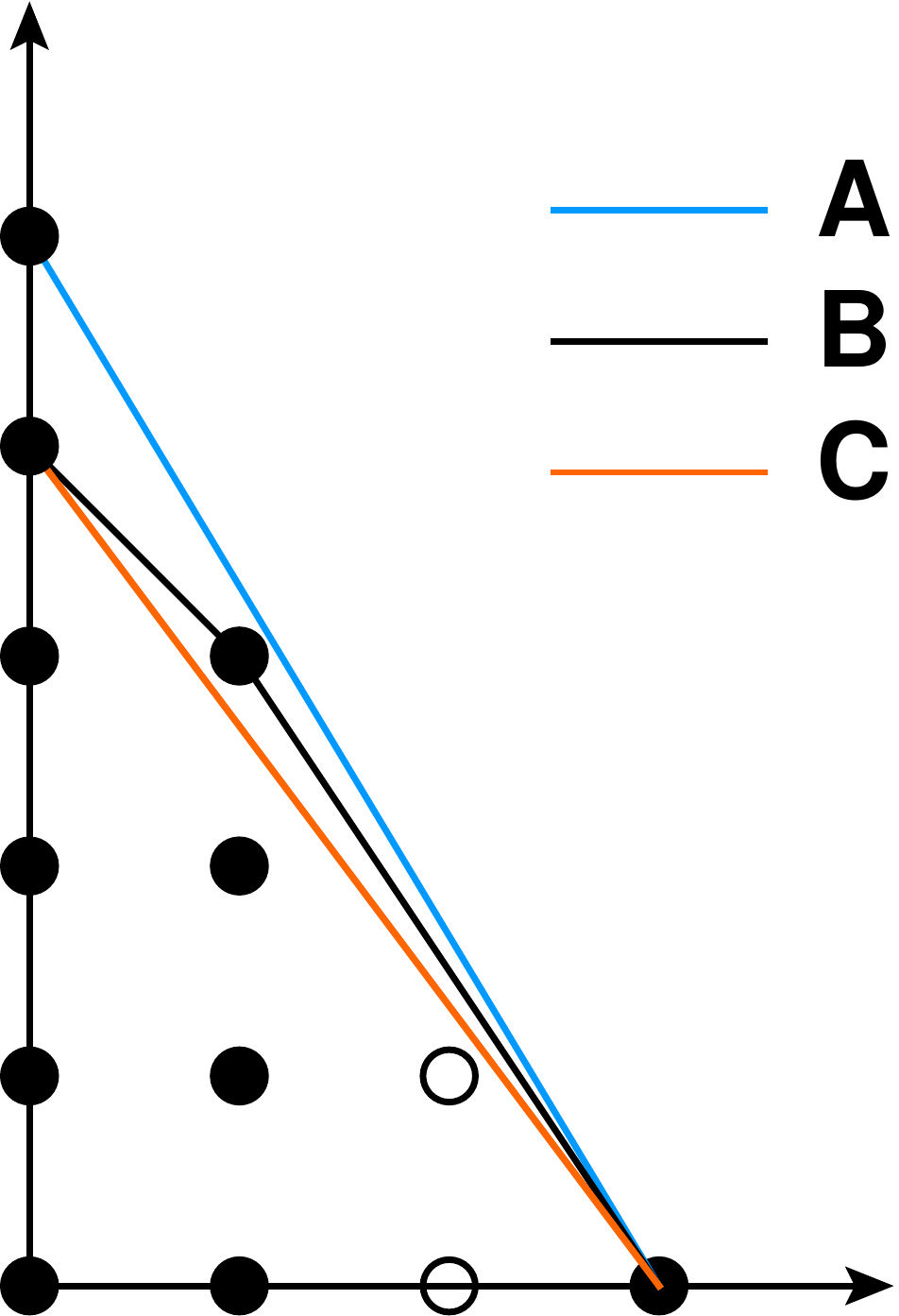}
\caption{The minimal flow from theory $A$ is to theory $B$ since the Newton polygon for $B$ is right below $A$. Similarly, the minimal flow from theory $B$ is to theory $C$.}
\label{RG}
\end{figure}
The above least RG flows can actually be seen easily from the Newton polygon: if the Newton polygon of theory $B$ is right below theory $A$, then the least flow is from $A$ to $B$. See figure \ref{RG}.

If there is an extra regular singularity to the type I singularity (we consider a simple singularity here since other cases can be flown to this one by higgsing the flavor symmetries), namely the $(\text{I}_{k,N},S)$ theory,
then the least relevant deformation is subtle and we can not naively perform the above kind of analysis. However, recall that this type IV theory has a realization as a type II theory which uses only one 
irregular singularity, and the Argyres-Douglas point is 
\eq{
x^{N+1}+xz^{k}=0.
}
The least relevant operator corresponds to turning on the deformation 
\eq{
x^{N+1}+xz^{k}+uz^{k+\left\lfloor{k\over N}\right\rfloor}=0,
}
and the new Argyres-Douglas point is $x^{N+1}+z^{k+\left\lfloor{k\over N}\right\rfloor}=0$. The regular singularity is gone since we have turned on the relevant operators from this singularity. So if $k<N$, the new IR fixed point of the minimal flow is actually the 
$\text{I}_{k,N+1}$ theory.

\begin{figure}
\center
\includegraphics[width=7.55cm]{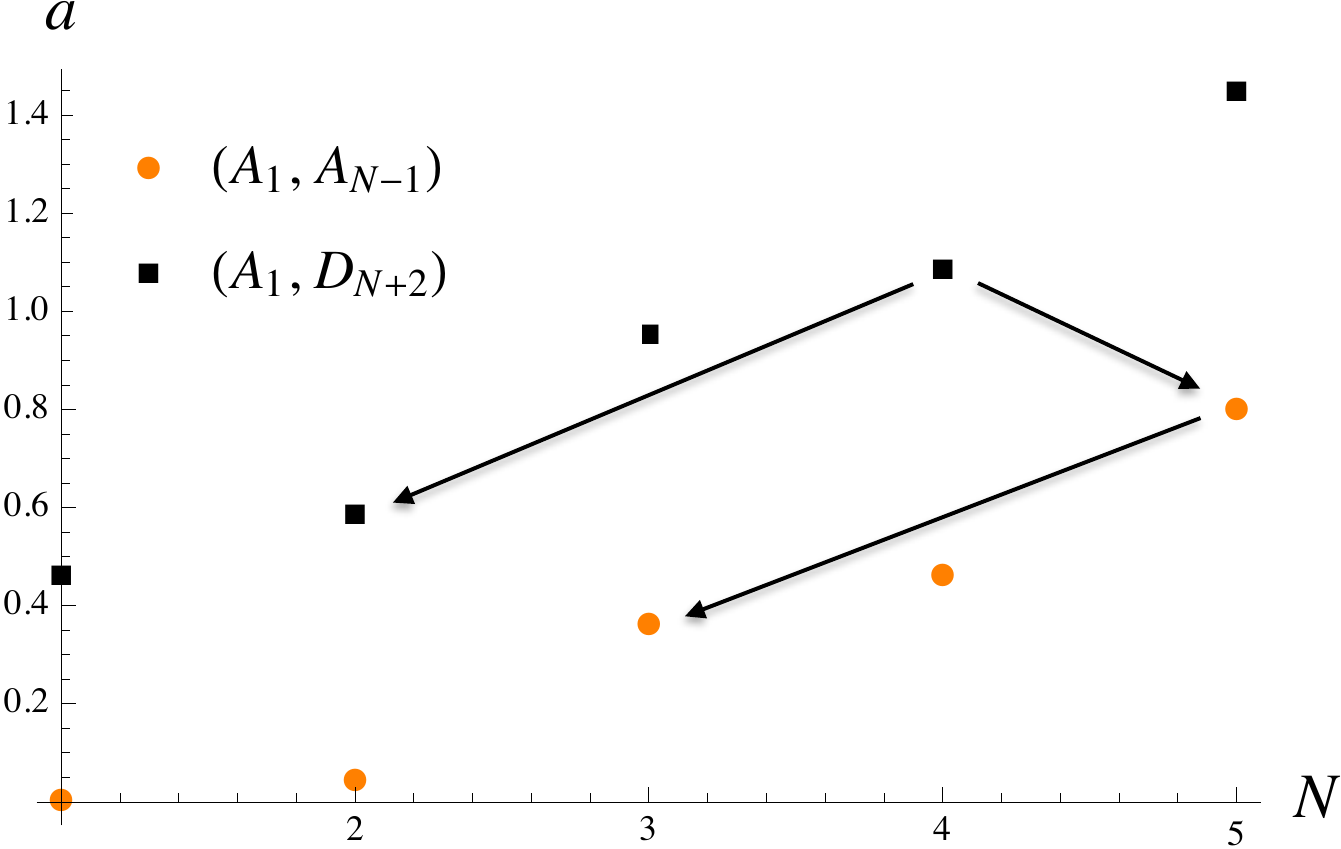}\quad\includegraphics[width=7.55cm]{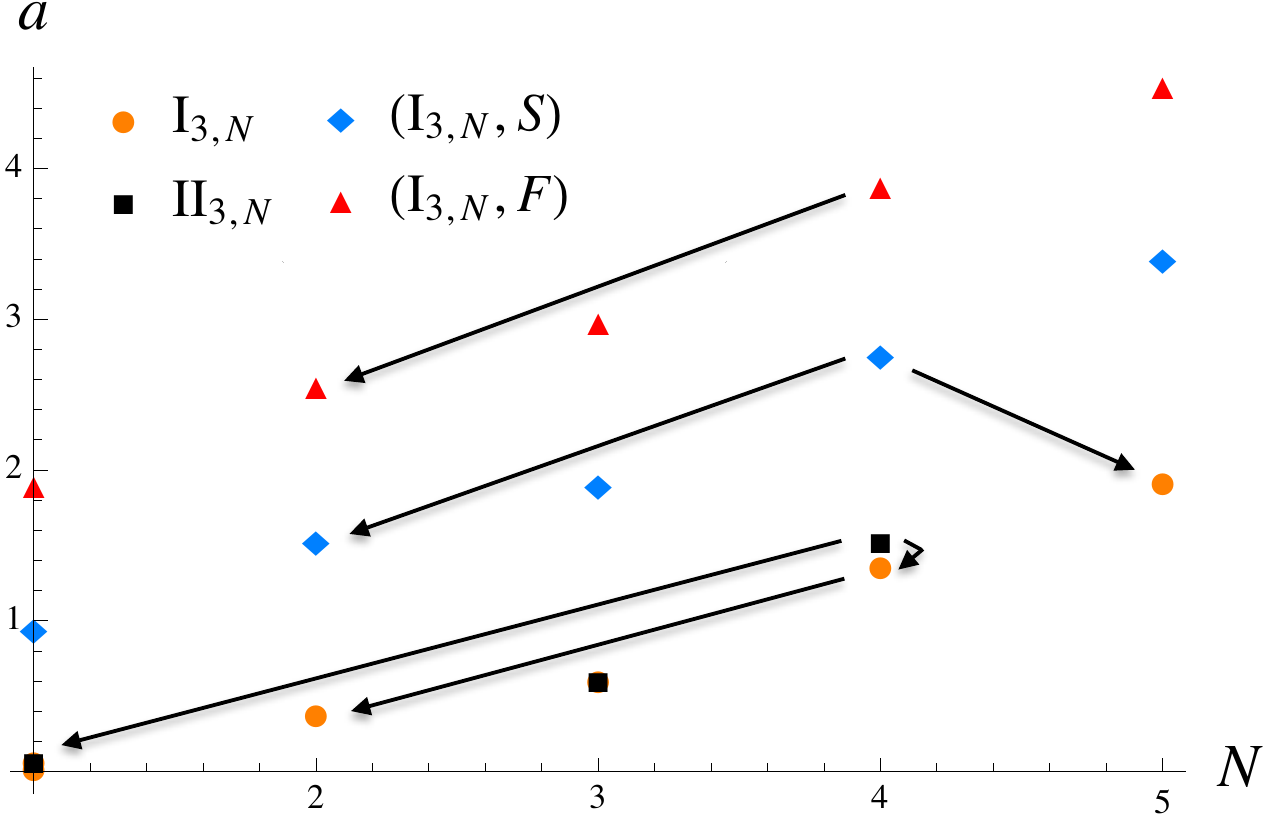}
\caption{The minimal flows between the $A_1$ and the $A_{2}$ Argyres-Douglas theories.}
\label{RG2}
\end{figure}
Since we have calculated the central charges for these SCFTs, let us check whether the $a$-theorem is obeyed for these flows. The RG flow patterns for the $A_{1}$ and the $A_{2}$ Argyres-Douglas theories are shown in figure \ref{RG2}. It is easy to check that all the RG flows satisfy the $a$-theorem using our explicit central charge formula.

\begin{figure}
\center
\includegraphics[width=15cm]{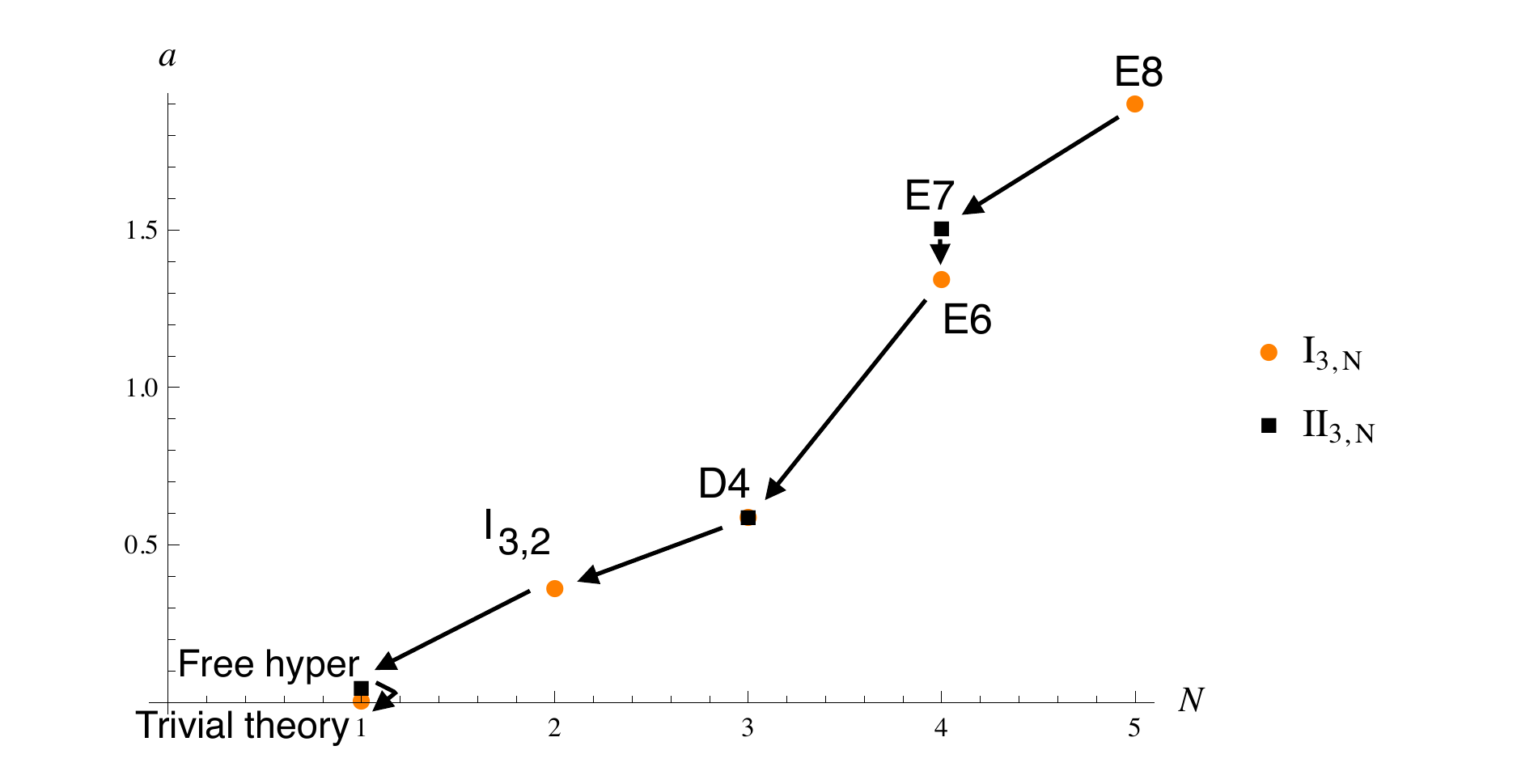}
\caption{RG flow from the $E_{8}$ Argyres-Douglas theory.}
\label{E8}
\end{figure}

\paragraph{Example 4:}
Let us study in detail the RG flow sequence by starting with the $E_8$ Argyres-Douglas theory.\footnote{The label in the $E_{n}$ ADE theory means that the corresponding BPS quiver has the $E_{n}$ shape. This should not be confused with the Minahan-Nemeschansky $E(n)$ theory, where $E(n)$ denotes the flavor symmetry \cite{Minahan:1996fg, Minahan:1996cj}.}
We begin with the $(A_{1}, E_{8}) \simeq (A_{2}, A_{4})$ curve at the Argyres-Douglas point
\eq{x^{3} + z^{5} = 0, \qquad [x] = \frac58, \quad [z] = \frac38, \qquad a = \frac{91}{48}, \quad c = \frac{23}{12}.}
We perturb by the term $uxz^{3}$ corresponding to the operator $u$ whose scaling dimension is $[u] = 1/8$. It becomes dominant in the IR and we obtain a new curve
\eq{x^{3} + xz^{3} = 0, \qquad [x] = \frac35, \quad [z] = \frac25, \qquad a = \frac{3}{2}, \quad c = \frac{31}{20}.}
Note that the dimensions of $x$ and $z$ have now changed. In the IR curve, the operator corresponding to the UV term $vz^{5}$ has dimension $[v] = -1/5$ and so is irrelevant.
The curve is the $(A_{1}, E_{7})$ curve. Let us keep flowing further into the IR. The relevant operator with the lowest dimension is $w z^{4}$, and we obtain the $(A_{1}, E_{6})\simeq (A_{2}, A_{3})$ curve
\eq{x^{3} + z^{4} = 0, \qquad [x] = \frac47, \quad [z] = \frac37, \qquad a = \frac{75}{56}, \quad c = \frac{19}{14}.}
The lowest possible dimension of the relevant operator in this theory is $2/7$, which is attained by the term $xz^{2}$. The SCFT is the $(A_{1}, D_{4})$ theory
\eq{x^{3} + xz^{2} = 0, 
\qquad a = \frac{7}{12}, \quad c=\frac{2}{3}.}
Flowing deeper into the IR, we obtain the $(A_{2}, A_{1})$ curve
\eq{x^{3} + z^{2} = 0, 
\qquad a = \frac{43}{120}, \quad c=\frac{11}{30}.}
Flowing even deeper into the IR, we obtain the $\text{II}_{3,1}$ curve
\eq{x^{3} + xz = 0, 
\qquad a=\frac{1}{24}, \quad c = \frac{1}{12}.}
This theory has only a single free hypermultiplet. Going all the way down, we obtain the $(A_{2}, A_{0})$ curve
\eq{x^{3} + z = 0,
\qquad a=0, \quad c=0.}
This is a trivial theory and the flow stops.
The RG flow pattern is shown in figure \ref{E8}.
\section{Flavor central charge $k_{G}$}\label{k}
The theories defined using the Riemann surface with defects in different duality frames can be thought of as gauging the flavor symmetry of two kinds of matter: a type IV Argyres-Douglas theory engineered using an irregular singularity and 
a regular singularity, and an isolated SCFT engineered using the sphere with three punctures. To calculate the contribution to 
the beta function of the gauge group, it is useful to find the central charge associated with the flavor group.

The central charge of the flavor group is defined as the coefficient $k_G$ of the leading term in the operator product expansion of two $G$-currents 
\eq{
J_\mu^a(x)J_\nu^b(0)={3k_G\over 4\pi^2}\delta^{ab}{x^2g_{\mu\nu}-2x_\mu x_\nu \over x^8}+\cdots
}
This normalization is taken such that $k_G=1$ for one fundamental matter of $SU(k)$. The $\mathcal{N}=2$ supersymmetry relates the current central charge $k_G$ 
to the 't Hooft anomaly via the relation
\eq{
k_{G}\delta^{AB}=-2~\!\text{tr}(RT^AT^B).
\label{flavor}
}

Our proposal for calculating $k_{G}$ for our class of theories is based on the following assumption: $k_G$ is equal to the dimension of the operator with the maximal scaling dimension among those operators from the regular singularity. 
Let us look at the $(A_1, D_{N+2})$ theory which is realized as a sphere with one irregular singularity and one regular singularity. Its Seiberg-Witten
curve is
\eq{
x^2=z^N+u_1z^{N-1}+\cdots+u_N+{u_{N+1}\over z}+{m^2\over z^2}.
} 
The central charge $k_{G}$ is given by the dimension of the operator $u_{N+1}$ coming from the regular singularity 
\eq{
k_G=[u_{N+1}]={2(N+1)\over N+2},
}
which is in exact agreement with the results in the literature \cite{Gaiotto:2010jf}. $k_{G}=1$ for $N=0$, which is good since the corresponding theory has just one hypermultiplet; $k_{G}=0$ for $N=-1$ which is also perfect since the theory is trivial.

This could be easily generalized to the higher-rank case. Let us take the type I irregular singularity that defines the $(A_{k-1}, A_{N-1})$ theory, and 
the regular singularity is the full one. The Seiberg-Witten curve is 
\eq{
x^k+\cdots+\left(z^N+v_1z^{N-1}+\cdots+{v_{N+1}\over z}+\cdots +{v_{N+k-1}\over z^{k-1}}+{v_{N+k}\over z^k}\right)=0.
}
Then the maximal scaling dimension of the local operators from the regular singularity is from $v_{N+k-1}$ and the central charge is
\eq{
k_G=[v_{N+k-1}]=k[x]+(k-1)[z]={k(N+k-1)\over N+k}.
}
When $N=-k+1$, we have $k_G=0$ which is correct since the theory is trivial.

If we use the irregular singularity corresponding to the type II Argyres-Douglas theory, then the Seiberg-Witten curve in the presence of the regular singularity is 
\eq{
x^k+\cdots+x(z^{N^{'}}+\cdots)+\left(v_0z^N+\cdots+{v_{N+k-1}\over z^{k-1}}+{v_{N+k}\over z^k}\right)=0.
}
The scaling dimensions are now $[x]={N^{'}\over N^{'}+k-1 }$ and $[z]={k-1\over N^{'}+k-1}$. $v_{N+k-1}$ is still the operator with the highest scaling dimension from the 
regular singularity. The central charge is 
\eq{
k_G=[v_{N+k-1}]={kN^{'}+(k-1)^2\over N^{'}+k-1}.
}
When $N^{'}=-k+2$, we have $k_G=1$ which is good since the theory has just one fundamental of $SU(k)$.

\section{Conclusion}\label{conclusion}
In this paper, we have calculated the central charges $a, c, k_G$ for a large class of superconformal field theories. They agree with all the previous calculations and 
include many new results. The RG flow between the general Argyres-Douglas theories can be described precisely and confirms the $a$-theorem. As we only did an elementary analysis, it would be nice to perform a more detailed study of the RG flow patterns.

Our calculation of the central charge $a$ suggests that there should be a good supergravity dual. It would be interesting to find the explicit gravity solution and to use other methods such as the holographic dual and the direct study of the Seiberg-Witten curve to confirm our calculation. It is straightforward to extend to superconformal field theories defined using the six-dimensional $D_N$ theory \cite{Tachikawa:2009rb} and regular singularities, and theories using other types of regular singularities \cite{Chacaltana:2012zy}. It would be interesting to calculate the central charges for other strongly-coupled theories considered in \cite{Argyres:2005pp, Argyres:2005wx, Argyres:2007tq, Cecotti:2011gu, DelZotto:2011an, Cecotti:2012jx, Cecotti:2010fi,Chacaltana:2012ch}.

The function $R(B)$ is related to the number of codimension one singularities on the Coulomb branch where extra massless dyons appear. These dyons should be included in the stable BPS spectrum which has been found in \cite{Xie:2012gd} for a large class of theories studied in this paper. It seems that $R(B)$ encodes the number of BPS particles in the maximal or minimal finite chamber. We have checked this for many examples which confirm the above conjecture, i.e. the $(A_1, A_{N-1})$ and $(A_1,D_{N+2})$ theory, where $R(B)$ gives the maximal chamber. 
Moreover, $R(B)$ for the $T_k$ theory (\ref{RBTk}) is equal to ${1\over 4} 2k(k-1)^2$ and this function reveals that one BPS chamber should have $2k(k-1)^2$ BPS states. Indeed, it is found in \cite{Xie:2012gd} that the minimal chamber of the $T_k$ theory has $2k(k-1)^2$ states. It would be interesting to further explore the relation between the function $R(B)$ and the BPS spectrum.
 
These Argyres-Douglas theories are like the minimal model of 4d $\mathcal{N}=2$ superconformal field theories. It would be worthwhile to explore other properties of them such as the conformal block, index, the relation to integrable model, etc. 
We believe that the study of these theories would provide many important insights into the understanding of the dynamics of quantum field theory. 

\appendix
\section{Review on topological gauge theory and the central charges}
In this appendix, we review the method proposed in \cite{Shapere:2008zf} for calculating the central charges. The $\mathcal{N}=2$ superconformal field theory has an
$SU(2)_R\times U(1)_R$ symmetry. If we introduce a background gauge field $W_{\mu\nu}^a$ for the $SU(2)_R$ symmetry, then the anomaly equation for the $U(1)_R$ current is
\eq{
\partial_{\mu}R^{\mu}={c-a\over8\pi^2} R_{\mu\nu\rho\sigma}\tilde{R}_{\mu\nu\rho\sigma}+{2a-c\over 8\pi^2}W_{\mu\nu}^aW^{\mu\nu}_a.
}

The topological twist of a four-dimensional $\mathcal{N}=2$ theory is done by setting the background gauge field equal to the self-dual part of the curvature of the four-manifold:
\eq{
W_{\mu\nu}^at_{\rho\sigma}^a={1\over2}(R_{\mu\nu\rho\sigma}+\tilde{R}_{\mu\nu\rho\sigma}).
}
After this twisting, one can find a nilpotent supercharge $Q$ and the theory makes sense on any curved manifold. By taking a proper basis for the $t$ matrix, we have
\eq{
W_{\mu\nu}^aW^{\mu\nu}_a={1\over 2}\left(R_{\mu\nu\rho\sigma}\tilde{\tilde{R}}_{\mu\nu\rho\sigma}+R_{\mu\nu\rho\sigma}\tilde{R}_{\mu\nu\rho\sigma}\right),
}
where $\tilde{\tilde{R}}_{\mu\nu\rho\sigma}={1\over 4}\epsilon_{\mu\nu}^{\ \ ab}\epsilon_{\rho\sigma}^{\ \ cd}R_{abcd}$.
The anomaly equation for the $\mathcal{N}=2$ $U(1)_R$ current of the twisted SCFT then becomes
\eq{
\partial_{\mu}R^{\mu}={2a-c\over 16\pi^2}R_{\mu\nu\rho\sigma}\tilde{\tilde{R}}_{\mu\nu\rho\sigma}+{c\over 16\pi^2}R_{\mu\nu\rho\sigma}\tilde{R}_{\mu\nu\rho\sigma}.
}
Similarly, one could introduce a background gauge field for the global symmetry $G$, and the integrated anomaly equation becomes
\eq{
\Delta R=2(2a-c)\chi+3c\sigma-k_G n,
}
where $\chi$ is the Euler characteristic, $\sigma$ is the signature of the four-manifold, and $n$ is the instanton number for the background gauge field 
of the global symmetry. A free hypermultiplet contributes $\Delta R= \sigma /4$ and a free vectormultiplet contributes $\Delta R= (\chi +\sigma )/2$ if there is no background flavor gauge field. 

Now we are going to use the 't Hooft anomaly matching condition to find the central charges from the low-energy description of the topologically twisted theory. The path integral for the IR theory 
has the following representation
\eq{
Z=\int [du][dq]A^\chi B^\sigma C^n e^{-S_\text{low-energy}}.
}
The measure factors $(A, B, C)$ can be used to compensate the IR $R$-anomaly. Let us first assume that there is no background gauge field for the flavor symmetry. If there are $r$ free vectormultiplets and $h$ free hypermultiplets at a generic point of the 
Coulomb branch, then the anomaly matching condition implies
\eq{
R(A)\chi+R(B)\sigma+\frac{r(\chi +\sigma )}{2}+ \frac{h\sigma}{4}=2(2a-c)\chi+3c\sigma.
}
Comparing the coefficients of $\chi$ and $\sigma$, we obtain two equations which we solve for the central charges in terms of the $R$-charges of the measure factors
\eq{a = \frac{R(A)}4 + \frac{R(B)}6 + \frac{5r}{24}+ \frac{h}{24}, \qquad c = \frac{R(B)}3 + \frac{r}6 + \frac{h}{12}.} 
The measure factors $A$ and $B$ are conjectured to take the following form
\eq{
A=\alpha\left[\det{\partial u_{i}\over \partial a^I}\right]^{\frac{1}{2}}, \qquad B=\beta\Delta^{^\frac{1}{8}},
}
where $\Delta$ is the physical discriminant which counts the number of singularities on the Coulomb branch. For this form, it is easy to see that
\eq{
R(A)=2\left[A\right]=\sum_i([u_i]-1), \qquad R(B) = \frac{1}{4}\left[\Delta\right],
}
and $\Delta$ is usually difficult to calculate.

\acknowledgments
We thank Anindya Dey, Chan Youn Park, Sanjaye Ramgoolam, Yuji Tachikawa for illuminating discussions, and Nick Dorey for comments on the manuscript. DX is supported in part by Zurich Financial services membership and by the U.S. Department of Energy, grant DE-FG02-90ER40542 (DX). 
PZ would like to thank the Institute for Advanced Study, the Carg\`ese summer school, the Yukawa Institute for Theoretical Physics, the Simons Center for Geometry and Physics, and the Perimeter Institute for Theoretical Physics for warm hospitality where this work was carried out. PZ is supported by a Dorothy Hodgkin Postgraduate Award from EPSRC and a Rouse Ball Traveling Studentship from Trinity College, Cambridge.

\end{document}